\documentclass[11pt, letterpaper]{article}
\usepackage{graphicx}
\usepackage[square,comma,numbers]{natbib}
\usepackage{amsmath}
\usepackage{amssymb}
\usepackage{amsmath}
\usepackage{amsfonts}
\usepackage{mathtools}
\usepackage[super]{nth}
\usepackage{setspace}
\usepackage{enumitem} 
\usepackage{epstopdf}
\usepackage{color}
\usepackage[letterpaper,left=1in,right=1in,top=1in,bottom=1in]{geometry}
\usepackage{algpseudocode}
\usepackage{algorithm}
\usepackage{multirow}
\usepackage{longtable}
\usepackage{rotating}
\usepackage{mathrsfs}
\usepackage{titlesec}
\usepackage{subcaption}
\usepackage{float}
\usepackage{graphicx}
\usepackage{booktabs}
\usepackage{caption}
\usepackage{bm}
\usepackage{tikz}
\usepackage{makecell}
\usepackage[toc,title,page]{appendix}
\usetikzlibrary{shapes,arrows,positioning,calc}

\titleformat{\paragraph}
{\normalfont\normalsize\bfseries}{\theparagraph}{1em}{}
\titlespacing*{\paragraph}
{0pt}{3.25ex plus 1ex minus .2ex}{1.5ex plus .2ex}

\title{\Large \bf A Semi-Centralized Multi-Agent RL Framework for  Efficient Irrigation Scheduling }
\author{
\centerline{\normalsize Bernard T. Agyeman$^{a}$, Benjamin Decardi-Nelson$^{b}$, Jinfeng Liu$^{a}$\thanks{Corresponding author: J. Liu. Tel: +1-780-492-1317. Fax: +1-780-492-2881. Email: jinfeng@ualberta.ca.}, Sirish L. Shah$^{a}$}
\vspace{2mm}\\
\centerline{\small $^{a}$Department of Chemical \& Materials Engineering, University of Alberta,}\\
\centerline{\small Edmonton, AB T6G 1H9, Canada.}\\ 
\centerline{\small $^{b}$Systems Engineering, Cornell University,}\\
\centerline{\small Ithaca, NY 14853, USA.}}
\allowdisplaybreaks

\begin{document}
\date{}
\maketitle

\setstretch{1.5}
{}
\begin{abstract}
This paper proposes a Semi-Centralized Multi-Agent Reinforcement Learning (SCMARL) approach for irrigation scheduling in spatially variable agricultural fields, where management zones address spatial variability. The SCMARL framework is hierarchical in nature, with a  centralized coordinator agent at the top level and decentralized local agents at the second level. The coordinator agent makes daily binary irrigation decisions based on field-wide conditions, which are communicated to the local agents. Local agents determine appropriate irrigation amounts for specific management zones using local conditions. The framework employs state augmentation approach to handle non-stationarity in the local agents' environments. An extensive evaluation on a large-scale field in Lethbridge, Canada, compares the SCMARL approach with a learning-based multi-agent model predictive control scheduling approach, highlighting its enhanced performance, resulting in water conservation and improved Irrigation Water Use Efficiency (IWUE). Notably, the proposed approach achieved a 4.0\% savings in irrigation water while enhancing the IWUE by 6.3\%.
\end{abstract}
\noindent{\bf Keywords}: Irrigation Scheduling, Mixed-Integer Optimal Control, Multi-Agent Reinforcement Learning, Semi-Centralized Multi-Agent Reinforcement Learning.  
\clearpage
\section{Introduction}
Agriculture is the leading consumer of freshwater worldwide, accounting for  70\% of global freshwater withdrawals~\cite{unwater2015}. The growing scarcity of freshwater, worsened by rapid population growth and climate change, continues to exert severe pressure on  freshwater resources. Although agriculture is significantly affected by this scarcity, it also contributes to the problem through excessive water use, primarily for irrigation. Furthermore, the water-use efficiency associated with current irrigation practices  remains inadequate, at about 60\%~\cite{world2009water}, highlighting the limitations of these practices. Implementing efficient water management strategies in agricultural irrigation will complement current efforts to mitigate the freshwater scarcity problem while ensuring the sustainability of the agricultural sector. One promising approach is closed-loop irrigation scheduling, which employs feedback to deliver precise amounts of water at optimal times. Closed-loop irrigation scheduling is evolving to include Management Zones (MZs)---defined as distinct areas within large-scale fields with uniform soil and crop properties. By integrating these MZs, closed-loop irrigation systems are better equipped to handle the significant soil variability in agricultural fields, further optimizing water utilization in irrigation.

In general, `scheduling' is the process of optimally allocating finite resources over a given time horizon to achieve specific objectives. In the domain of irrigation, scheduling involves determining the optimal water quantities and timing for irrigation. This process typically occurs on an hourly or daily basis. In the daily irrigation scheduling problem, which is the focus of this study, the objective is to identify the specific days (irrigation time) within a scheduling horizon for irrigation, as well as the precise amounts of water (irrigation amount) to be applied during each event. Furthermore, in agricultural fields characterized by several MZs, the daily irrigation scheduling problem extends to determining a uniform irrigation timing that applies across all MZs, while simultaneously specifying the optimal irrigation amounts tailored to the specific needs of each Management Zone (MZ).

Model Predictive Control (MPC), an optimal control method, has been extensively used to solve scheduling problems~\cite{subramanian2012state, yi2015adaptive,subramanian2014economic}.
MPC seeks to determine control actions that optimize a cost function over a fixed horizon. In MPC, control actions are calculated by solving an optimization problem that is constrained by a process model.
The suitability of MPC for scheduling is enhanced by the ability to model scheduling problems in state-space form~\cite{subramanian2012state} and by the adaptation of the MPC cost function to include measurable economic metrics~\cite{risbeck2018closed}. Specifically, in the field of irrigation scheduling, techniques such as set-point tracking MPC and zone tracking MPC with continuous-valued controls have been employed to optimize irrigation schedules~\cite{delgoda2016irrigation, mccarthy2014simulation, nahar2019closed}.

Scheduling problems inherently involve combinatorial aspects, necessitating discrete decisions within the scheduling frameworks to ensure optimal allocation of limited resources. For example in the daily irrigation scheduling problem, deciding on which days to irrigate within the scheduling horizon reduces to making discrete, binary decisions. In this process, each day of the scheduling horizon is assigned a binary (0/1 or `yes/no') decision variable. A value of 1 assigned to a binary variable signifies that irrigation should take place on that specific day, while a value of 0 indicates that no irrigation should occur on that day. Consequently, scheduling models typically incorporate both continuous and discrete controls, with mixed-integer linear and nonlinear programming (MILP and MINLP) methods widely employed to model these problems. Recent research has demonstrated the feasibility of integrating discrete-valued controls directly into the MPC framework with minimal impact on stability properties~\cite{rawlings2017model}. This integration has led to the adoption of mixed-integer MPC to address scheduling challenges in various domains, including chemical production systems~\cite{risbeck2018closed,subramanian2012state}. In the field of irrigation scheduling, a new approach, namely LSTM-based mixed-integer MPC with zone control, was proposed in \cite{agyeman2023lstm} to address the daily irrigation scheduling problem.

Even the simplest scheduling problems are classified as NP-hard, meaning that the direct application of  MILP/MINLP solvers to scheduling models is only practical for small-scale problems~\cite{sundaramoorthy2005simpler}. 
Mixed-integer MPC problems that have been employed to solve irrigation scheduling problems naturally inherit this complexity, and this complexity becomes  pronounced when applied to fields comprising of several MZs. 
The necessity of obtaining irrigation schedules within a reasonable time-frame has prompted the development of approximation methods for mixed-integer MPC-based irrigation schedulers. One such method involved utilizing the logistic sigmoid function to model the binary decision variables within the resulting MINLP formulation, which transformed the MINLP to a nonlinear program~\cite{agyeman2023lstm}. While this approach enables the calculation of irrigation schedules within an acceptable time-frame, it is prone to approximation errors and presents difficulties related to the interpretability of its results. 

Across various domains, reinforcement learning (RL) has emerged as a practical approach to solving complex problems. In RL, an agent is trained to make sequential decisions that maximize cumulative rewards. The emergence of approaches such as hierarchical structures~\cite{wei2018hierarchical} and hybrid actor-critic methods~\cite{fan2019hybrid}, which adapt RL algorithms to both continuous and discrete action spaces, has broadened the applicability of RL to mixed-integer optimal control problems. In the context of daily irrigation scheduling, a centralized agent capable of handling both discrete and continuous action spaces offers a straightforward method for determining the daily `yes/no' irrigation decision and the corresponding irrigation amounts across MZs.
 While this centralized approach simplifies coordination, it also presents several practical challenges. It is prone to a single point of failure and suffers from scalability issues, making it less practical for large-scale implementations.

One potential way to address the scalability issue associated with a centralized agent is to deploy Multi-Agent Reinforcement Learning (MARL), potentially in a decentralized fashion. In a MARL setting, multiple agents, each with distinct observations and actions, collaborate to make collective decisions. MARL offers benefits compared to single-agent RL approaches, including: 1) improved efficiency by decomposing complex problems into simpler sub-problems that can be distributed across agents, 2) enhanced robustness, as failures in single agents can be compensated for by other agents, and 3) improved scalability.

A recent study~\cite{agyeman2024learning} employed decentralized RL agents, along with a heuristic approach and decentralized MPC to address the daily irrigation scheduling problem. This study focused on training hybrid Proximal Policy Optimization (PPO) agents independently for each MZ, where each agent determined the daily irrigation decision and the irrigation amount.  
The need for uniform irrigation decisions across all zones, as required by the daily irrigation scheduling problem, posed a challenge within the decentralized framework. To address this, the study adopted a heuristic approach to establish a uniform irrigation decision that was enforced across all MZs. Subsequently, a decentralized MPC method, integrating the uniform irrigation decision, was employed to calculate the irrigation amounts for each MZ. While this method was found to be computationally efficient and outperformed the conventional irrigation scheduling technique in terms of Irrigation Water Use Efficiency (IWUE), it faced limitations due to the sub-optimal nature of the approach used to determine the uniform irrigation decision. Although the decentralized approach was efficient at determining irrigation amounts, which are local in nature, it was not effective in ensuring an optimal field-wide irrigation decision.  The lack of field-wide optimality arises because the decentralized agents make irrigation decisions based on local observations, and the heuristic method used to unify these irrigation decisions does not guarantee an optimal solution for the entire field. Furthermore, the decentralized agent framework, coupled with the heuristic approach, necessitated the solution of the decentralized MPC to determine the irrigation amounts across MZs, which comes at a computational cost. 

An alternative, potentially more robust approach, would be to integrate the strengths of both centralized and decentralized RL agents. Drawing parallels from the approaches utilized in the design of coordinator MPC~\cite{aske2008coordinator} could prove valuable in this context. Coordinator MPC employs a semi-centralized control approach, dividing control responsibilities between local and centralized controllers. 
Similarly, exploring semi-centralized MARL for irrigation scheduling in fields with multiple MZs, where the `yes/no' daily irrigation decision and the daily irrigation rates across various MZs are respectively distributed between a centralized RL agent and fully decentralized RL agents, presents a promising approach. Semi-centralized multi-agent reinforcement learning has demonstrated its effectiveness in various domains. For instance, a semi-centralized deep deterministic policy gradient algorithm was proposed for cooperative tasks in StarCraft games~\cite{xie2020semicentralized}, and a semi-centralized multi-agent reinforcement learning algorithm involving PPO and deep Q-networks was introduced to maximize energy efficiency in Internet-of-Things networks~\cite{alajmi2022semi}.  

MARL can be described as the application of RL within Multi-Agent Systems (MASs). It is noteworthy that MASs have found application in irrigation scheduling~\cite{jimenez2020survey}. For example, agent-based modeling, a development within MASs, has been widely used for decision management in irrigation systems~\cite{belaqziz2011agent,wanyama2017multi}. It is important to mention that the agents in these MASs were programmed with predefined behaviors. However, due to the complexity of most environments and the limitations of pre-programmed agent behaviors, it has been observed that learning new behaviors online is often necessary to improve the performance of agents and the MAS as a  whole. This is particularly important in dynamic environments such as agricultural fields, where fixed agent behaviors may become inappropriate. RL enables autonomous agents to learn new behaviors online, making it particularly attractive for application in MAS. While MARL has been employed  in~\cite{agyeman2024learning} to address the irrigation scheduling problem in a complementary manner, the full potential of RL in learning agent behavior within the context of MAS in irrigation scheduling remains unexplored.

The main contribution of this work is to explore the full potential of RL in learning agent behavior within the context of MAS in irrigation scheduling by proposed a Semi-Centralized MARL (SCMARL) framework to address the daily irrigation scheduling problem in spatially variable fields that are characterized by several MZs. The proposed framework is hierarchical in nature, with a coordinator agent at the top of the framework. This coordinator agent is a centralized RL agent with a discrete action space, and its task is to determine the daily `yes/no' irrigation decision based on  a field-wide soil moisture content, weather data, and crop information. At the lower level of the SCMARL framework are local agents, one for each MZ in the field. These local agents operate in a decentralized manner and are tasked with determining the optimal irrigation amounts for their respective MZs. Their decisions are based on the irrigation decision made by the coordinator agent, as well as local soil moisture content, crop data, and weather information. Furthermore, the proposed framework employs a state augmentation approach, which involves the integration of the irrigation decision into the input of a local agent's policy, to address the issue of non-stationarity that arises in MARL settings.

This paper extends the findings presented in~\cite{agyeman2024semi}. In contrast to~\cite{agyeman2024semi}, the current work provides detailed explanations of the SCMARL algorithm and framework. Additionally, this study introduces a simulation experiment that assesses the benefits of the state augmentation approach in stabilizing learning in the proposed framework. Furthermore, the paper conducts a more comprehensive evaluation of the SCMARL framework in a large-scale setting, comparing it to a learning-based MPC approach that employs MARL and MPC in a complementary manner to address the irrigation scheduling problem.

\section{Preliminaries}
The concept of a Markov Decision Process (MDP), which is relevant to the single-agent scenario, is first  introduced.\\ 
{Definition 1}: A fully-observable MDP is represented as $\langle S, A, p, r \rangle$, where $S$ signifies the set of states in the environment, $A$ denotes the available actions for the agent, $p : S \times A \times S \rightarrow [0, 1]$ is the probability distribution governing state transitions, and $r : S \times A \times S \rightarrow R$ is the reward function.
In state $s_k$ with action $a_k$ taken, the agent transitions to state $s_{k+1}$ with probability $p(s_k, a_k, s_{k+1})$. Correspondingly, the agent receives a reward $r(s_k, a_k, s_{k+1})$ as feedback from the environment. The agent's goal is to find an optimal policy $\pi: S \rightarrow A~(\text{or}~\pi : S \times A \rightarrow [0, 1]$ for stochastic policies) to maximize the expected discounted return.

Next, the concept of a stochastic game (SG), or Markov game, which extends the MDP framework to the multi-agent setting is defined as:\\
Definition 2:  A fully-observable SG is represented as $\langle N, S_i, A_i, p, r_i \rangle$, where  $i\in N$ and $N = [1, . . . , n]$ is the set of $n$ agents. $S_i$ is the set of states, $A_i$ represents the possible actions for n agents, leading to the joint action set as the Cartesian product of the action sets for each agent, i.e., $A = A_1 \times A_2 \times A_3 ... A_n$. Similar to the single-agent case, $p : S \times A \times S \rightarrow [0, 1]$ is the state transition probability distribution, and $r_i : S \times A \times S \rightarrow R, \forall i \in N$ are the reward functions for agents.

\section{Problem Formulation}\label{sec:problem_formulation}
This work addresses the daily irrigation scheduling problem for a large-scale agricultural field delineated into distinct irrigation MZs. The objective is to develop  scheduler that determines both the daily irrigation decisions  (i.e., `yes' or `no'), applied uniformly across all MZs, and the specific irrigation application amounts for each zone. The problem is formulated as follows:

\textbf{Given:}
\begin{itemize}
	\item \textbf{Scheduling Horizon}: The known number of days for which the irrigation needs to be scheduled.
	\item \textbf{Management Zones ($M$)}: The field is divided into $M$ distinct MZs.
	\item \textbf{Weather Predictions}: Daily predictions of reference evapotranspiration and precipitation, essential for determining the water needs of crops. Reference evapotranspiration estimates the water lost through evaporation and plant transpiration, while precipitation serves as a natural water input to soil.
	\item \textbf{Crop Information}: This includes the crop coefficient, derived from empirical relations calibrated specifically for the crop and field under study. The crop coefficient adjusts the daily reference evapotranspiration predictions to reflect the specific water needs of the crop at different growth stages, ensuring that irrigation schedules meet the precise water requirements of the crops.
	\item \textbf{Soil Moisture Content}: Initial distribution of soil moisture content in the rooting depth across each MZ at the start of the scheduling horizon. This information is essential for determining the existing water availability and the required irrigation to maintain optimal soil moisture levels for crop growth.
\end{itemize}
	
	Determine:
\begin{itemize}
	\item \textbf{Irrigation Decision}:  A binary `yes/no' decision for irrigation for each day of the scheduling horizon. In accordance with standard irrigation practice, the irrigation decision should apply uniformly across all MZs of the field, and as such  must be optimal from a field-wide point of view.
	\item \textbf{Irrigation Amounts}: The daily irrigation amount for each MZ zone for every day within the scheduling horizon. The scheduler should ensure that the irrigation amounts align with the binary irrigation decision---prescribing zero irrigation amounts for `no' days and a non-zero irrigation amounts for `yes' days.
\end{itemize}

\section{System Description}
\begin{figure}[H]
	\centering
	\includegraphics[width=1\columnwidth]{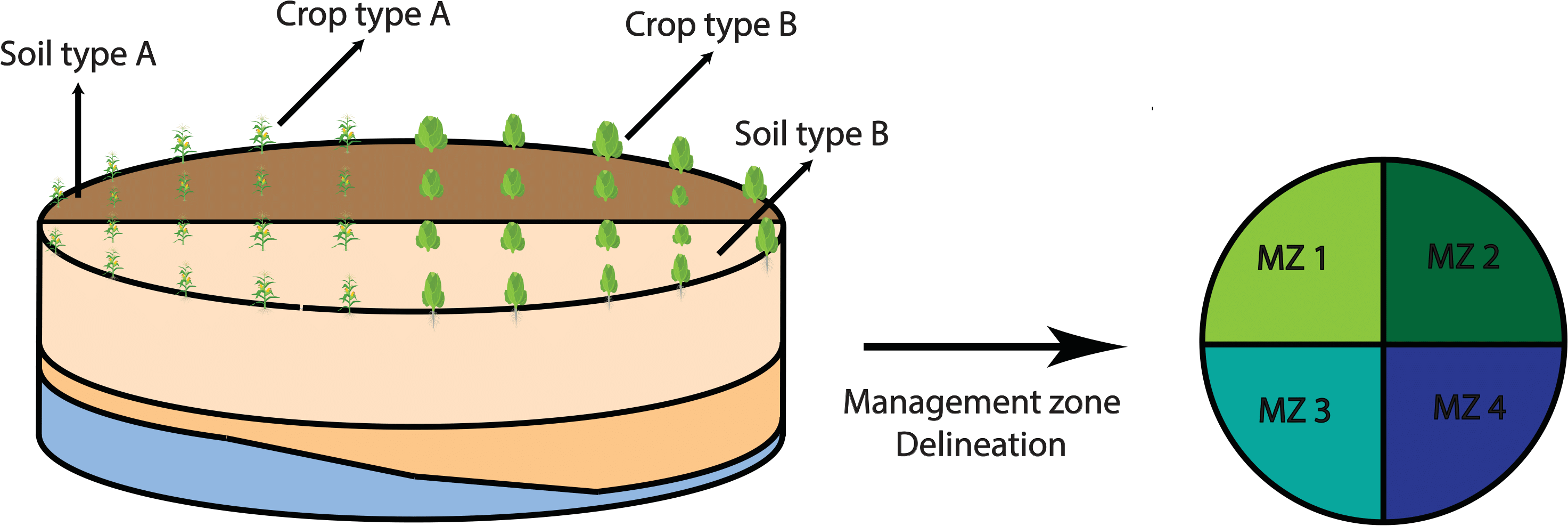}
	\caption{A diagrammatic representation of a field with variability in crop and soil types. The field is divided into 4 distinct MZs, each with uniform soil and crop properties.}
	\label{fig:spat_var_field}
\end{figure}
The system under study is a large-scale, spatially-variable agro-hydrological field divided into distinct irrigation MZs. A MZ is defined as a sub-field with uniform soil and crop properties. Figure~\ref{fig:spat_var_field} illustrates a field that exhibits spatial variability in terms of soil type and crop type. Given that the field comprises 2 soil types and 2 crop types, and according to the definition of a MZ, the field is delineated into 4 distinct MZs.

While various attributes can be employed for MZ delineation, the proposed framework is best suited for a delineation approach that considers key attributes such as elevation and soil hydraulic parameters. Elevation affects the movement and distribution of water across agricultural fields, thereby influencing irrigation efficiency. Similarly, soil hydraulic parameters (which can be inferred from soil texture or soil electrical conductivity) directly impact soil moisture dynamics and water availability to crops, which are essential for effective irrigation scheduling.

The use of MZs to address spatial variability in a large-scale agricultural field allows for an independent approach to model soil moisture dynamics in each zone, due to the uniformity in crop and soil type that exists in the various MZs.
This independence is particularly feasible in large-scale fields, where generally weak interactions between the MZs minimize the impact of adjacent zones on each other’s water dynamics. 

In this work, water transport within each MZ is modeled using the Richards equation, a mechanistic agro-hydrological model that accounts for various hydrological processes, including irrigation, rain, evapotranspiration, transpiration, infiltration, root-water extraction, surface runoff, and drainage.

Specifically, the 1D version of the Richards equation, parameterized by the soil hydraulic parameters specific to the uniform soil type within each MZ, is employed to describe soil moisture dynamics. The selection of the 1D model is justified by proposing the use of elevation data in the delineation process, which ensures relatively flat elevation profiles within each MZ. Consequently, soil water transport in each MZ is expected to be predominantly influenced by vertical (axial) dynamics rather than lateral movements. The 1D version of the Richards equation is expressed as:
\begin{equation}\label{eq:RE_1D}
	C(\psi)\frac{\partial \psi}{\partial t} = \frac{\partial}{\partial z}\bigg[K(\psi)\bigg(\frac{\partial \psi  }{\partial z}+ 1\bigg)\bigg]-\rho(\psi)\mathcal{R}\left(\text{K}_{\text{c}}, \text{ET}_0, \text{z}_{\text{r}}\right)
\end{equation}
In Equation~(\ref{eq:RE_1D}), $\psi~(m)$ is the capillary pressure head, which describes the status of water in soil, $t~(s)$ represents time, $z~(m)$ is the spatial coordinate, $K(\cdot)~(m \cdot s^{-1})$ is the unsaturated hydraulic water conductivity and  $ C(\cdot)~(m^{-1})$ is the capillary capacity. 
$\rho(\cdot)~(-)$ is a dimensionless stress water factor, $\mathcal{R}(\cdot)$ is the root water uptake model which is a function of the crop coefficient $\text{K}_{\text{c}}~(-)$, the reference evapotranspiration $\text{ET}_0 ~(m \cdot s^{-1})$, and the rooting depth $\text{z}_{\text{r}}~(m)$. Interested readers may refer to~\cite{agyeman2021soil}
for a detailed description of $K(\cdot)$, $C(h)$, $\rho(\cdot)$ and $\mathcal{R}$.

To solve the 1D Richards equation numerically, the following boundary conditions are typically imposed:
\begin{alignat}{2}
	&\frac{\partial (\psi+z)}{\partial z}\bigg|_{z=H_z}&&=1\label{eq:botBC} \\
	&\frac{\partial \psi}{\partial z}\bigg|_{z=0}&&=-1-\frac{u^{\text{irr } } - \text{EV}}{K(\psi)}\label{eq:topBC}
\end{alignat}
where $H_z~(m)$, $u^{\text{irr}}~(m \cdot s^{-1})$, and $\text{EV}~(m\cdot s^{-1})$ in Equations (\ref{eq:botBC}) and (\ref{eq:topBC})  represent the depth of the soil column, the irrigation amount and the evaporation rate, respectively. Once a numerical value of the capillary pressure head $\psi$ is obtained, the  corresponding volumetric soil moisture content $\theta_v$ can be calculated using the following relationship:
\begin{equation}\label{eq:thetareln}
	\theta_{{v}} (\psi)=\theta_r +(\theta_{{s}}-\theta_{{r}})\bigg[\frac{1}{1+(-\alpha \psi)^n}\bigg]^{1-\frac{1}{n}}
\end{equation}
where $\theta_s$ is the saturated moisture content, $\theta_{{r}}$ is the residual moisture content, and $\alpha$ and $n$ are curve fitting parameters used to describe the soil water retention curve.  These parameters, including the saturated hydraulic  conductivity $K_s$  are the relevant soil hydraulic parameters required to solve the Richards equation. Collectively, these 5 parameters are denoted in this work as  $\bm{\phi} = \left[ K_s, \theta_s, \theta_r, \alpha, n \right]$.

The  parameterized 1D Richards equation, after carrying out the temporal and spatial discretizations can be written in state-space form as:
\begin{align}
	x_{k+1}&=\mathcal{F}(x_k,u_k, \bm{{\phi}}) + \omega_k \label{eq:state_equation}\\
	y_{k}&=\mathcal{H}(x_k,\bm{{\phi}}) + v_k \label{eq:output_equation}
\end{align}
where $x_k\in \mathbb{R}^{N_x}$ represents the state vector containing $N_x$ capillary pressure head values for the spatial nodes in the soil column. In Equations~(\ref{eq:state_equation}) and~(\ref{eq:output_equation}), $\mathcal{F}$ and $\mathcal{H}$ represent the system dynamics and output function, respectively. $u_k$ represents the input vector containing the irrigation amount, precipitation, daily reference evapotranspiration, the crop coefficient, and the rooting depth. The terms $\omega_k$ and $v_k$ represent the uncertainties in the state and output equations, respectively. The volumetric water content $\theta_{\text{v}}$ is chosen as the output $y_k$. Equation~(\ref{eq:output_equation}) is thus a general representation of Equation~(\ref{eq:thetareln}) and $y_k\in \mathbb{R}^{N_y}$ represents the output vector containing $N_y = N_x$ volumetric soil moisture content values for the corresponding spatial nodes in the soil column.

Based on the independent modeling approach and the application of the 1D Richards equation to capture soil moisture dynamics in each MZ,  for a field with $M$ MZs,  the state vector ${\bm{X}}_k \in \mathbb{R}^{N_x\times M} $ and the output vector ${\bm{Y}}_k \in\mathbb{R}^{N_y\times M} $  can be compactly represented as follows:

\begin{equation}\label{eq:state_delineated_field}
{\bm{X}}_{k+1} = 
\begin{bmatrix}
{x}_{k+1, 1}\\
{x}_{k+1, 2}\\
 \vdots \\
 {x}_{k+1, M}\\

\end{bmatrix}  =  \begin{bmatrix}
\mathcal{F}\left({x}_{k, 1}, u_{k,1}, \bm{\phi}_1\right) + \omega_{k,1} \\
\mathcal{F}\left({x}_{k, 2}, u_{k,2}, \bm{\phi}_2\right)+ \omega_{k,2}\\
 \vdots \\
 \mathcal{F}\left({x}_{k, M}, u_{k,M}, \bm{\phi}_M\right) +\omega_{k,M}\\
\end{bmatrix}
\end{equation}

\begin{equation}\label{eq:output_delineated_field}
{\bm{Y}}_{k} = 
\begin{bmatrix}
{y}_{k, 1}\\
{y}_{k, 2}\\
 \vdots \\
 {y}_{k, M}\\

\end{bmatrix}  =  \begin{bmatrix}
\mathcal{H}({x}_{k, 1}, \bm{\phi}_1) + v_{k,1} \\
\mathcal{H}({x}_{k, 2}, \bm{\phi}_2)+ v_{k,2}\\
 \vdots \\
 \mathcal{H}({x}_{k, M},\bm{\phi}_M) + v_{k,M}\\
\end{bmatrix} 
\end{equation}
In Equations~(\ref{eq:state_delineated_field}) and~(\ref{eq:output_delineated_field}), $\bm{\phi}_i$ represents the set of hydraulic parameters specific to the uniform soil type in MZ $i$, used to parameterize $\mathcal{F}$ and $\mathcal{H}$. Additionally, $x_{k, i}$, $u_{k,i}$ and $y_{k,i}$ represent the state, input and output vectors of MZ $i$ at time instant $k$, while $\omega_{k,i}$ and $v_{k,i}$ represent the uncertainties in the state and output equations, respectively. 

\section{Proposed Approach: Semi-Centralized MARL}
\begin{figure}
	\centering
	\includegraphics[width=0.8\columnwidth]{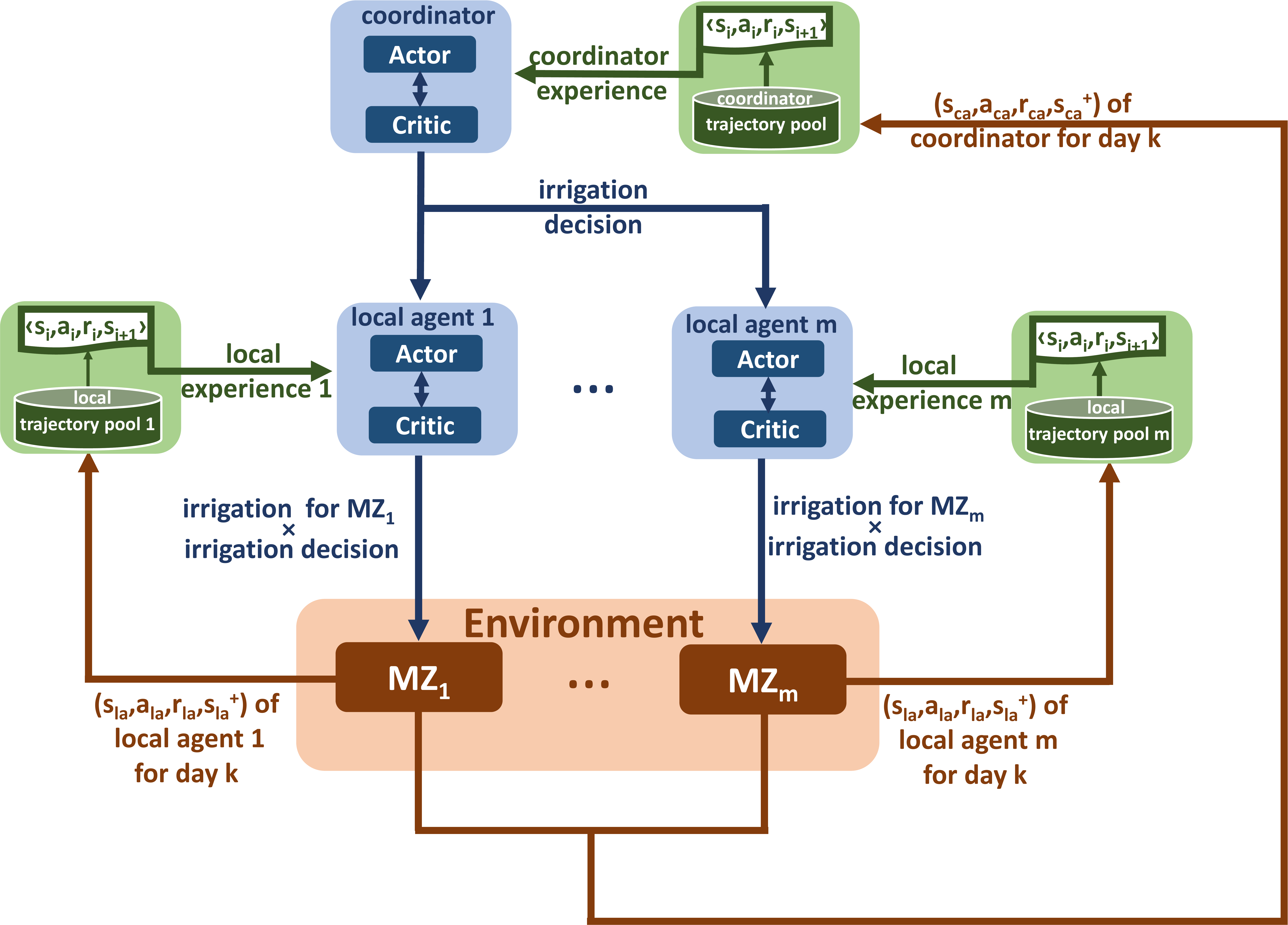}
	\caption{A schematic diagram of the Semi-Centralized MARL framework for irrigation scheduling.}
	\label{fig:schd_fd}
\end{figure}

To effectively manage  daily irrigation scheduling  in a large-scale agricultural field  composed of multiple MZs, this study proposes  a two-tier hierarchical  semi-centralized MARL approach, depicted in  Figure~\ref{fig:schd_fd}. The framework consists of a coordinator agent  and local agents, each responsible for different aspects of the scheduling problem.

At the highest level of the hierarchy is the coordinator, which serves as the root/central node.  This agent, which assumes the role of a centralized agent in the framework, is responsible for determining  the daily binary irrigation decision (`yes/1' or `no/0'). The coordinator considers soil moisture information gathered from all MZs of the field along with weather and crop information during the determination of the irrigation decision. By accounting for soil moisture information from all MZs, the coordinator agent  is expected to prescribe an optimal `yes/no' irrigation decision for the entire field. This decision acts as a master control, directly enabling or disabling the irrigation amounts recommended by the local agents, ensuring adherence of the entire framework to the  daily irrigation strategy outlined in Section~\ref{sec:problem_formulation}.

Local agents, which function as decentralized agents in the framework, are responsible for recommending the daily irrigation amounts for the MZs based on local soil moisture information, weather, crop information and the coordinator's irrigation decision. The irrigation amounts proposed by local agents are adjusted by the coordinator's decision: multiplied by `1' for a `yes/1' decision leading to 	irrigation, or by `0' for a `no/0' decision resulting in no irrigation. 

%

While any RL algorithm can be employed to learn agent behavior in the proposed framework, this study proposes the use of an actor-critic RL  algorithm to train all agents in the SCMARL framework. Specifically, the PPO algorithm~\cite{schulman2017proximal} is utilized to learn agent behavior within the SCMARL framework. PPO is a policy gradient method designed to optimize an agent’s policy by estimating both the value function (critic) and the policy (actor). This actor-critic framework uses the value function to estimate expected returns and guide policy updates through the estimation of a metric known as the \textit{advantage}, which measures how much better an action is compared to the average action at a given state.

During the training of the agents, PPO collects experience data by having the agent interact with the environment and uses this data to update both the policy and value function. The policy update is performed by minimizing a surrogate loss function, which approximates the policy’s performance in terms of the expected reward. This update process is guided by the estimated advantages provided by the value function. PPO further enhances stability in the learning process by using a clipped objective function, which restricts policy updates to prevent large deviations from the current policy that could destabilize the learning process. Given its robustness and ability to guarantee stable policies, PPO is well-suited for the environments considered in the proposed SCMARL framework.

As can be seen  in Figure~\ref{fig:schd_fd}, each agent within the SCMARL framework maintains its own trajectory pool. This pool is used for storing experiences, which are essential for updating the agent's actor and critic networks. The proposed SCMARL method for irrigation scheduling in large-scale fields characterized by multiple MZs  is summarized in Algorithm~\ref{alg:scmarl}.

\subsection{Non-stationarity in the SCMARL Framework} 

In MARL settings, where multiple agents concurrently learn and update their policies, the transition dynamics and rewards from the perspective of any single agent are inherently non-stationary. This non-stationarity arises because an agent's environment is influenced not only by its own actions but also by the joint actions of all agents in the system. In the literature, approaches such as centralized training and
decentralized execution, sequential iterative best
response and multi-timescale learning have been proposed to address the non-stationarity problem within MARL settings~\cite{nekoei2023dealing}.

In the SCMARL framework designed for large-scale agricultural fields, each MZ is modeled independently, facilitated by the weak coupling between MZs. This design significantly reduces dependencies between local agents, minimizing the impact of one local agent’s actions on another’s reward and state transition dynamics. 
However, the coordinator agent introduces a central source of non-stationarity in the learning environments of the local agents by making a binary decision (`yes/1' or `no/0') that controls irrigation across the field. This decision fundamentally alters the local agents' potential actions and their learning environments by directly impacting the transition dynamics and potential rewards.

To address this non-stationarity in the learning environments of the local agent, the framework employs two main strategies:
\begin{itemize}
	\item \textbf{Communication of  Irrigation Decision}: Local agents are informed of the coordinator’s decision before making their irrigation amount  recommendations. 
	
	\item \textbf{State Augmentation}: The coordinator's decision, combined with local soil moisture, weather, and crop information, is integrated through an augmentation process. This augmented state then serves as the input for the policy of a local agent.
	
\end{itemize}
This dual approach stabilizes the learning environment for local agents by ensuring they are aware of the global irrigation decision. 
This reduces the uncertainty in the local agents' learning environments by informing them of the irrigation decision that will ultimately be applied. Additionally, the inclusion of the coordinator’s decision in the local agents' policy inputs and the switch-like implementation of this decision during training encourage the local agents to align their irrigation recommendations with the coordinator’s decisions. They learn to prescribe non-zero irrigation amounts when a `yes/1' decision is made and zero irrigation amounts for `no' decisions, based on the rewards and feedback from their environments. This learning outcome reduces redundancy during the execution phase of the SCMARL framework by minimizing the need to explicitly enforce the coordinator's decision and enhances the autonomy of the agents.

It is important to note that non-stationarity is less pronounced from the perspective of the coordinator agent, compared to the local agents. The high-level decisions made by the coordinator directly influence the final actions of the local agents. This direct influence means that the coordinator agent imposes a consistent strategy on the actions of the local agents, thereby reducing variability in its environment.

\subsection{Local Agent Design}
For each MZ, a local agent is assigned with the primary purpose of determining the daily optimal irrigation amount. Subsequent sections will detail the essential MDP elements that govern the decision-making framework for a local agent within the SCMARL framework.

\subsubsection{Transition Dynamics}
During the training of local agents within the proposed SCMARL framework, the environmental dynamics are modeled using a well-calibrated 1D Richards equation. Calibration of this equation is essential for an accurate description of the soil moisture dynamics, and it involves parameterizing Equations~(\ref{eq:state_equation}) and~(\ref{eq:output_equation}) with soil hydraulic parameters $\bm{\phi}$, which are representative of the uniform soil type found across the MZ. Although this manuscript does not address the process of determining these parameters, data assimilation techniques, such as those proposed in \cite{agyeman2022simultaneous}, could be utilized to estimate these parameter based on real-time soil moisture observations.

The environmental dynamics are detailed in Equations~(\ref{eq:state_equation}) and (\ref{eq:output_equation}). These equations update the soil water content within each MZ based on inputs such as the local agent's irrigation amount, the initial soil moisture content, and the prevailing weather and crop conditions.

\subsubsection{State Design}
Each local agent in the framework receives an input vector ${y}$, which represents the spatial volumetric soil moisture contents derived from Equation (\ref{eq:output_equation}). This vector is combined with other relevant data for irrigation scheduling: daily reference evapotranspiration ($\text{ET}_{0}$), daily crop coefficient ($\text{K}_{\text{c}}$), and rainfall/precipitation ($\text{R}_{\text{n}}$). Additionally, each agent receives the daily irrigation decision ($c$) made by the coordinator. Thus, the state of a local agent can be compactly represented as $s_{\text{la}} \coloneqq [{y}, \text{ET}_{0}, \text{K}_{\text{c}}, \text{R}_{\text{n}}, c]$.

The actor network $\mathcal{A}_{\text{la}}$ of a local agent takes $s_{\text{la}}$ as input and outputs the daily irrigation amount $a_{\text{la}}$ for its MZ. This amount is then adjusted by the coordinator's decision $c$, resulting in the final irrigation action $u^{\text{irrig}}_{\text{la}} \coloneqq a_{\text{la}} \times c$, which is applied to the environment.

Following the application of $u^{\text{irrig}}_{\text{la}}$, the soil moisture content ${y}$ transitions to the successor soil moisture content ${y}^{+}$, according to Equations~(\ref{eq:state_equation}) and~(\ref{eq:output_equation}). To determine the successor state $s^{+}_{\text{la}}$, the one-day-ahead weather and crop coefficient predictions are utilized, yielding updated conditions $\text{ET}^{+}_{0}$, $\text{K}^{+}_{\text{c}}$, and $\text{R}^{+}_{\text{n}}$. Simultaneously, the coordinator's updated actor network $\mathcal{A}_{\text{ca}}$ is employed to predict the next day's irrigation decision $c^{+}$ using the aggregated information from all MZs, which includes ${y}^{+}$ for each MZ and the forecasted weather data. This aggregated state is represented as $s^{+}_{\text{ca}} \coloneqq [{y}^+_1, {y}^{+}_2, ..., {y}^{+}_M, \text{ET}^{+}_0, \text{K}^{+}_{\text{c}}, \text{R}^{+}_{\text{n}}]$, from which $c^{+} = \mathcal{A}_{\text{ca}}(s^{+}_{\text{ca}})$ is derived. Consequently, the updated state for each local agent is $s^{+}_{\text{la}} \coloneqq [{y}^+, \text{ET}^{+}_0, \text{K}^{+}_{\text{c}}, \text{R}^{+}_{\text{n}}, c^{+}]$.

\subsubsection{ Reward Design}\label{sec:reward_design}
Each local agent within the framework is tasked with determining the daily irrigation amount, denoted as $a_{\text{la}}$, for its assigned MZ.  The primary objective is to maintain the daily root zone soil moisture content $\theta^{\text{RZ}}$ within a target soil moisture range, bounded by an upper limit $\overline{\nu}$ and a lower limit $\underline{\nu}$. $\bar{\nu}$ and $\underline{\nu}$ (also known as the threshold volumetric moisture content $\theta_{\text{th}}$) are calculated as follows:
\begin{align}\label{eq:zone_bounds}
	\bar{\nu} &= \theta_{\text{fc}}\\
	\underline{\nu} &=\theta_{\text{th}} =  \theta_{\text{fc}} - \left[ \text{MAD}\times(\theta_{\text{fc}} -  \theta_{\text{wp}})\right]
\end{align}
where $\theta_{\text{fc}}$ is the volumetric moisture content at field capacity, $\theta_{\text{wp}}$ represents the volumetric water content at the wilting point, and MAD refers to the management allowable depletion, which indicates the fraction of the total available water that is permitted to be depleted. 

The root zone soil moisture content $\theta^{\text{RZ}}$ is calculated as a weighted sum of the spatial soil moisture content ${y}$,  with 40\% of the weight assigned to the average moisture in the top quarter of $\text{z}_{\text{r}}$, 30\% to the average moisture in the second quarter of $\text{z}_{\text{r}}$, 20\% to the average moisture in the third quarter of $\text{z}_{\text{r}}$, and 10\% to the average moisture in the bottom quarter of $\text{z}_{\text{r}}$.

In addition to maintaining moisture within $\overline{\nu}$ and $\underline{\nu}$, the local agent aims to minimize the daily irrigation amount as a secondary objective, thereby conserving water. The reward function $r_{\text{la}}$ of a local agent consists of two parts: the target range tracking reward $r_{\text{la}}^z$ and the irrigation amount minimization reward $r_{\text{la}}^u$:
\begin{equation}\label{eq:rew_local_agent}
	r_{\text{la}} \coloneqq \alpha_{\text{la}}r_{\text{la}}^z + \beta_{\text{la}}r_{\text{la}}^u
\end{equation}
where $\alpha_{\text{la}}$  and $\beta_{\text{la}}$ are weights for the target range tracking and irrigation amount minimization rewards, respectively.
The target range tracking reward $r_{\text{la}}^z$ is defined as:  
\begin{align}
	r^{\text{z}}_{\text{la}} &\coloneqq \begin{cases}
		-\underline{Q}\times|\theta^{\text{RZ}} - \underline{\nu}| & \text{if } \theta^{\text{RZ}} < \underline{\nu}\\
		-\overline{Q}\times|\theta^{\text{RZ}} - \overline{\nu}| & \text{if } \theta^{\text{RZ}} > \overline{\nu}\\ 
		0 & \text{if } \underline{\nu} \leq \theta^{\text{RZ}} \leq \overline{\nu}\\
	\end{cases} \label{eq:rew_la}
\end{align}
with $\underline{Q}>0$, $\overline{Q}>0$ as adjustable weights that penalize deviations from the target moisture range. Note that $\theta^{\text{RZ}}$ is calculated with the successor soil moisture content $y^+$, which is the updated spatial soil moisture content resulting from the application of the modified version of the local agent's action to its respective MZ. The local agent's action is modified by multiplying its prescription with the coordinator's decision before it is applied to its environment/MZ. 

From Equation~(\ref{eq:rew_la}), it can be seen that  the local agent incurs a penalty of $-\underline{Q}\times|\theta^{\text{RZ}}_k - \underline{\nu}|$ when its actions result in the root zone soil moisture content falling below the lower threshold $\underline{\nu}$. Similarly, the local agent is penalized with $-\overline{Q}\times|\theta^{\text{RZ}}_k - \overline{\nu}|$ if its actions cause the root zone soil moisture content to exceed the upper threshold $\overline{\nu}$. When the local agent successfully maintains the root zone soil moisture within the desired range, it receives a reward of 0.

The irrigation amount minimization reward $r_{\text{la}}^u$ is defined as:
\begin{equation}
	r_{\text{la}}^u \coloneqq -R_u a_{\text{la}}
\end{equation}
where $R_u>0$ represents the per unit cost of the water used, reflecting the economic impact of water usage.

\subsection{Coordinator Agent Design}
The function of the coordinator agent within the framework is to make the daily irrigation decision ($c$), which is binary `yes/no'. For each MZ of the field, the decision of the coordinator agent ensures that its root zone soil moisture content lies within a predefined soil moisture range. 
Subsequent sections will detail the essential elements of the coordinator agent's configuration.
\subsubsection{Transition Dynamics}

During the training stage of the coordinator agent, $M$ independently calibrated 1D Richards equations for the $M$ MZs within the field are employed to model the environmental dynamics. These equations are compactly represented by Equations~(\ref{eq:state_delineated_field}) and~(\ref{eq:output_delineated_field}). The calibration process involves using the $M$ hydraulic parameters that accurately represent the soil type present in each of the $M$ MZs. For the simulations of these dynamics, the initial soil moisture contents are drawn from the respective MZs within the field.

It is important to mention that these 1D Richards equations are predefined during the training of the $M$ local agents. Therefore, there is no necessity to explicitly redefine these $M$ 1D Richards equations as individual components during the training of the coordinator agent. The coordinator agent can thus rely on the same set of $M$ 1D Richards equations that are employed in the training of the local agents.

\subsubsection{State Design}

The coordinator agent receives as input a concatenation of the $M$ spatial volumetric moisture contents, represented as $Y\coloneqq[{y}_{1},{y}_{2},\dots,{y}_{M}]$. Alongside $Y$, the inputs include daily reference evapotranspiration ($\text{ET}_{0}$), the crop coefficient ($\text{K}_{\text{c}}$), and precipitation ($\text{R}_{\text{n}}$), forming the state of the coordinator agent, denoted as $s_{\text{ca}} \coloneqq [Y,\text{ET}_{0},\text{K}_{\text{c}},\text{R}_{\text{n}}]$.

The coordinator’s actor network, $\mathcal{A}_{\text{ca}}$, processes $s_{\text{ca}}$ to determine the daily irrigation decision $c$ applicable across the entire field. This decision effectively activates (`on') or deactivates (`off') the irrigation amounts determined by local agents for each MZ. Following the application of of the joint action $ A \coloneqq [c\times a_{\text{la}_1}, c\times a_{\text{la}_2} \dots, c\times a_{\text{la}_M} ]$ of the coordinator and local agents  to the entire field, the concatenated spatial volumetric water content $Y$ transitions to ${Y}^{+}$, according Equations~(\ref{eq:state_delineated_field}) and~(\ref{eq:output_delineated_field}). The next state, $s^{+}_{\text{ca}}$ incorporates ${Y}^{+}$  along with forecasts for the following day’s weather conditions ($\text{ET}^{+}_{0},\text{K}^{+}_{\text{c}},\text{R}^{+}_{\text{n}}$) and it compactly represented as $[{Y}^{+}, \text{ET}^{+}_0, \text{K}^{+}_{\text{c}}, \text{R}^{+}_{\text{n}}]$.
  
\subsubsection{Reward Design}
Primarily, the coordinator agent's objective is to determine the daily irrigation decision $c$ that ensures the root zone soil moisture content in each MZ ($\theta^{\text{RZ}}_i~\forall i \in [1,2,\dots,M]$)  remains within a predetermined target range. Additionally, it aims to minimize the fixed costs associated with irrigation operations.
Consequently, the reward $r_{\text{ca}}$ of a local agent consists of two main parts: the target range tracking reward $r_{\text{ca}}^z$ across all $M$ MZs and the fixed irrigation cost minimization reward $r_{\text{ca}}^c$:
\begin{equation}\label{eq:rew_coordinator_agent}
	r_{\text{ca}} \coloneqq \alpha_{\text{ca}}r_{\text{ca}}^z + \beta_{\text{ca}}r_{\text{ca}}^c
\end{equation}
where $\alpha_{\text{ca}}$  and $\beta_{\text{ca}}$ are weights for the target range tracking in all the $M$ MZs and  fixed irrigation cost minimization rewards, respectively.

The target range tracking reward, $r_{\text{ca}}^z$, is calculated as the summation of individual rewards for each MZ:
\begin{equation}
	r_{\text{ca}}^z \coloneqq \sum_{i=1}^{M} r^{\text{z}}_{\text{la},i}
\end{equation}  
where $r^{\text{z}}_{\text{la}}$ is calculated using Equation~(\ref{eq:rew_la}) for a particular MZ.

The  fixed irrigation cost minimization reward, $r_{\text{ca}}^c$,   reflects the cost implications of the irrigation decision:
\begin{equation}
	r_{\text{ca}}^c \coloneqq -R_cc
\end{equation}
where $R_c$  is a positive weight that quantifies the fixed cost of performing the irrigation event, which seeks to promote cost effective irrigation practices.

\begin{algorithm}
\small
	\caption{SCMARL  for Irrigation Scheduling in Spatially Variable Fields.}\label{alg:scmarl}
	\begin{algorithmic}[1]
		\State Initialize weights of critic $\mathcal{C}_{\text{ca}}$ and actor $\mathcal{A}_{\text{ca}}$ networks of the coordinator agent.
		\For {local agent $i\gets 1$ to $M$}
		\State Initialize weights of critic $\mathcal{C}_{\text{la}_i}$ networks of $i$.
		\State Initialize weights of actor $\mathcal{A}_{\text{la}_i}$ networks of $i$.
		\EndFor
		\State Set the total number of episodes $K$.
		\State Set the horizon length for each episode $T$.
		\For {episode $k\gets 1$ to $K$}
		\For {local agent $i\gets 1$ to $M$}
		\State Obtain initial soil moisture (${y}_{i,\text{init}}$) of $i$.
		\State Set ${y}_i \gets {y}_{i,\text{init}} $.
		\EndFor
		\State Obtain initial weather and crop coefficient data $\text{ET}_{0,\text{init}}$, $\text{R}_{\text{n},\text{init}}$,~$\text{K}_{c,\text{init}}$.
		\State Set  $\text{ET}_{0} \gets \text{ET}_{0,\text{init}}$, $\text{R}_{\text{n}} \gets \text{R}_{\text{n},\text{init}}$,~$\text{K}_{c} \gets \text{K}_{c,\text{init}}$.
		\For {timestep $t\gets 1$ to $T$}
		\State Obtain $s_{\text{ca}}=[{y}_1,{y_2},...,{y}_M,\text{ET}_0,\text{K}_{\text{c}},\text{R}_{\text{n}}]$.
		\State Select $a_{\text{ca}}=c = \mathcal{A}_{\text{ca}}(s_{\text{ca}})$.
		\For {local agent $i\gets 1$ to $M$}
		\State Obtain  $s_{\text{la}_i} = [{y}_i,\text{ET}_0,\text{K}_{\text{c}},\text{R}_{\text{n}},c]$.
		\State Obtain  $a_{\text{la}_i} =  \mathcal{A}_{\text{la}_i}(s_{\text{la}_i})$ for agent $i$.
		\State Execute $u^{\text{irr}}_{\text{la}_i} =a_{\text{la}_i} \times c$ in environment of $i$.
		\State Obtain  $y_i^{+}$ from Equations~(\ref{eq:state_equation}) and~(\ref{eq:output_equation}).
		\State Obtain $r_{\text{la}_i}$ from Equation~(\ref{eq:rew_local_agent}).
		\EndFor
		
		\State Store states $[s_{\text{la}_1},s_{\text{la}_2},\dots,s_{\text{la}_M}]$.
		\State Store actions  $[a_{\text{la}_1},a_{\text{la}_2},\dots,a_{\text{la}_M}]$.
		\State Store rewards  $[r_{\text{la}_1},r_{\text{la}_2},\dots,r_{\text{la}_M}]$.
		\State Store next states  $[y^+_{\text{la}_1},y^+_{\text{la}_2},\dots,y^+_{\text{la}_M}]$.

		\State Obtain weather  and crop coefficient predictions as $\text{ET}^{+}_{0}$, $\text{R}^{+}_{\text{n}}$,~$\text{K}^{+}_{c}$.
		\State Obtain $s^{+}_{\text{ca}}=[{y}^+_1,{y}^{+}_2,\dots,{y}^{+}_M,\text{ET}^{+}_0,\text{K}^{+}_{\text{c}},\text{R}^{+}_{\text{n}}]$.
		\State Obtain $r_{\text{ca}}$ from Equation~(\ref{eq:rew_coordinator_agent}).
		\State Store ($s_{\text{ca}},a_{\text{ca}},r_{\text{ca}},s^{+}_{\text{ca}}$) for coordinator.
		\State Update $\mathcal{C}_{\text{ca}}$.
		\State Update $\mathcal{A}_{\text{ca}}$.
		\State Obtain $c^{+} = \mathcal{A}_{\text{ca}}(s^{+}_{\text{ca}})$.

		\For {local agent $i\gets 1$ to $M$} 
		\State Obtain $s^{+}_{\text{la}_i}=[{y}^{+}_i,\text{ET}^{+}_0,\text{K}^{+}_{\text{c}},\text{R}^{+}_{\text{n}},c^{+}]$.
		\State Store ($s_{\text{la}_i},a_{\text{la}_i},r_{\text{la}_i},s^{+}_{\text{la}_i}$) for $i$.
		\State Update $\mathcal{C}_{\text{la}_i}$  for $i$.
		\State Update $\mathcal{A}_{\text{la}_i}$  for $i$.
		\EndFor
		\For {local agent $i\gets 1$ to $M$}
		\State Set ${y}_{i} \gets {y}^+_{i}$.
		\EndFor
		\State Set  $\text{ET}_{0} \gets \text{ET}^{+}_{0}$, $\text{R}_{\text{n}} \gets \text{R}^{+}_{\text{n}}$,~$\text{K}_{c} \gets \text{K}^{+}_{c}$.
		\EndFor
		\EndFor
	\end{algorithmic}
\end{algorithm}

\normalsize
\section{Experimental Setup and Design}\label{sec:experimental_setup}

\begin{figure}[!ht]
	\subfloat[Study Area]{
		\begin{minipage}[c]{0.50\columnwidth}
			\centering
			\includegraphics[width=\columnwidth]{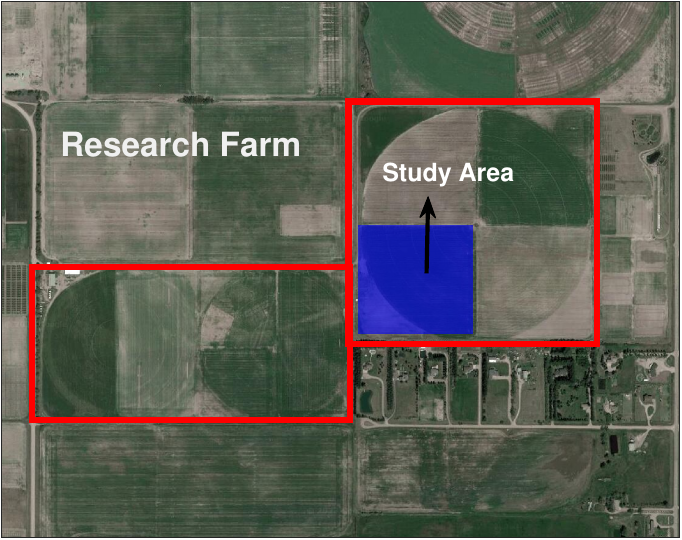}
	\end{minipage}}
	\hfill
	\hspace{4mm}
	\subfloat[Management Zone Map.]{
		\begin{minipage}[c]{0.47\columnwidth}
			\centering
			\includegraphics[width=\columnwidth]{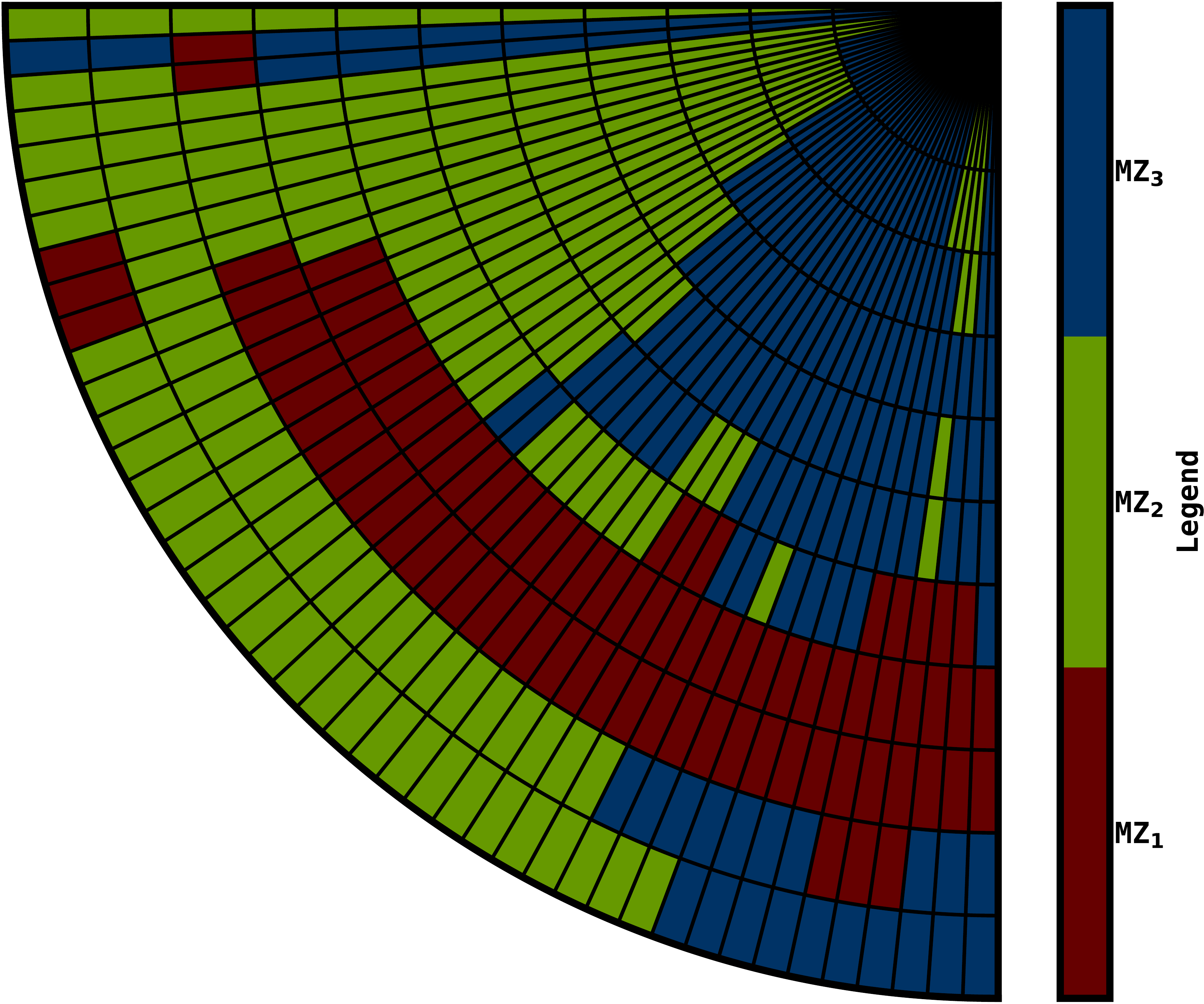}
	\end{minipage}}
	\caption{Study area and its management zone map.}
	\label{fig:study_area_mzd}
\end{figure}
The proposed SCMARL framework was applied to schedule irrigation within a specific  quadrant of a large-scale circular field, highlighted by the blue rectangle in Figure~\ref{fig:study_area_mzd}(a). This study area is located at a research farm in Lethbridge, Alberta, Canada, with geographic coordinates of 49.72° N and 112.80° W, and is managed by Lethbridge College. The field is equipped with a variable rate center pivot irrigation system, spanning a length of 294 meters, to facilitate  irrigation management. 

Prior to the application of the SCMARL approach, the investigated quadrant was delineated into MZs using a three-stage delineation method originally proposed in \cite{agyeman2024learning}. This method incorporated elevation and soil hydraulic parameters attributes, and it employed the k-means clustering technique for MZ delineation. The soil hydraulic parameter attributes were obtained from an offline data assimilation method, which assimilated remotely sensed soil moisture observations into the cylindrical coordinates version of the Richards equation using the extended Kalman filtering technique. This approach effectively estimated both soil moisture content  and the soil hydraulic parameters within the quadrant. Figure~\ref{fig:study_area_mzd}(b) illustrates the three distinct MZs that were identified in the investigated quadrant.

As each MZ corresponds to a cluster, the hydraulic parameters at the cluster's centroid are expected to be representative of the entire MZ's hydraulic parameters. These centroidal hydraulic parameters can thus be employed to calibrate the 1D Richards equations used to model the soil moisture dynamics in the study area. Given that the hydraulic parameters used for MZ delineation are estimated based on soil moisture observations from the field, the calibrated 1D Richards equations are expected to accurately describe the soil moisture dynamics in the field under study.

Additionally, the experimental setup included the cultivation of wheat within the study area. For this experiment, four agents were trained: three local agents corresponding to the three MZs, and one coordinator agent.

\subsection{Environment Setup and Design}
Each of the calibrated 1D Richards equations considered a soil column with a depth of 0.50 meters, discretized into 21 equally spaced compartments. Spatial discretization was performed using the central difference scheme, while backward differentiation was employed for temporal discretization.  Although the transition from the current soil moisture content to the next, under the prescribed irrigation amounts, considered a total time corresponding to 1 day (1440 minutes), a 30-minute step size was employed to minimize truncation errors in the temporal domain.

In each of the three MZs, the spatial soil moisture contents were initialized with converged soil moisture estimates obtained from the offline soil moisture and soil hydraulic parameter estimation process detailed in Section~\ref{sec:experimental_setup}. Note that using these converged soil moisture estimates to initialize the spatial soil moisture contents in the three MZs informed the training of the agents within the MDP framework. Daily reference evapotranspiration data were uniformly generated between 1.04 mm/day and 9.0 mm/day, reflecting the range of historical evapotranspiration observed in the study area. Historical rainfall data from the 2005 to 2020 growing seasons were also incorporated. Additionally, crop coefficient values for wheat were calculated using an empirical relation based on historical daily mean temperature from 2005 to 2020 in the study area. Details of this empirical relation can be found in Appendix~\ref{sec:crop_coeff}. A rooting depth of 0.50 m was employed, consistent with common practices for irrigation scheduling in wheat cultivation.

To account for discrepancies between the actual field and the calibrated 1D Richards equations, and to create more robust and generalizable policies, noise was included in the environmental models used. Specifically, the terms $\omega$ and $v$ in Equations~(\ref{eq:state_equation}) and~ (\ref{eq:output_equation})  for each MZ were sampled from normal distributions with zero means and standard deviations of 0.0002 and 0.0005, respectively. These values were derived from experience with using the Richards equation to simulate soil moisture dynamics in agricultural fields. Additionally, to reflect the typical imperfections in weather and crop information, noise was added to the daily weather and crop data during the training of the agents. The noisy versions of the weather and crop data were used by the agents, while the actual values were employed in the simulation of the Richards equations.

\subsection{Agent Configuration and Training}\label{sec:training_scmarl}
In the  design of  the reward functions for the agents, the reward parameters $\alpha_{\text{la}}$ and $\beta_{\text{la}}$  were set to 1.0 for the local agents, and $\alpha_{\text{ca}} = 0.1$ and $\beta_{\text{ca}} = 1.0$ for the coordinator agent. Furthermore, $\bar{Q}$ and $\underline{Q}$   were set at 1200000 and  1000000,  respectively, with $R_c = 1000$, and $R_u = 9000$. Notably, the reward parameters employed in the experimental study were treated as tuning parameters. Through an iterative process, the performance of the agents was evaluated under different configurations of these values, leading to the determination of the most suitable values for achieving the desired irrigation objectives.

The MAD value, relevant for calculating $\bar{\nu}$ and $\underline{\nu}$, was set at 50\%. For $\text{MZ}_1$ and $\text{MZ}2$, $\bar{\nu}$ and $\underline{\nu}$ were set at 0.280 and 0.200, respectively. For $\text{MZ}3$, the values for $\bar{\nu}$ and $\underline{\nu}$ were set as 0.30 and 0.230, respectively. The specific values of $\theta_{\text{fc}}$ and $\theta_{\text{wp}}$ which were used to determine $\bar{\nu}$ and $\underline{\nu}$ for the MZs were obtained from Reference~\cite{huffman2012irrigation}.

The actor and critic networks for each agent were trained with a learning rate of $1 \times 10^{-5}$.  Inputs to the networks  were normalized using the min-max aproach, where the inputs were bounded between -2.0 and 2.0. Other relevant   agent training settings included a time horizon of 30, a minibatch size of 64, 20 epochs, a discount factor of 0.99, a generalized advantage estimation parameter of 0.97, a clipping parameter of 0.25, and an entropy coefficient of 0.01. 

The coordinator agent’s decisions, given that it is discrete, were generated using a softmax distribution,  and each local agent calculates its daily irrigation rate through a Gaussian distribution. The policies of the agents were represented with  a fully connected multi-layer perceptron with two hidden layers, each containing 64 neurons and using a hyperbolic tangent activation function. The environment setup assigns 25 inputs to each local agent and 66 inputs to the coordinator agent, in accordance with the proposed SCMARL approach for irrigation scheduling in spatially-variable fields. The agents were configured and trained using the Tensorforce library~\cite{schaarschmidt2018lift} in Python, for 9,800 episodes over 10 runs.

\section{Performance Evaluation}
Three main studies were employed to assess the performance of the proposed SCMARL framework. In the first study, the benefits of integrating decisions from the coordinator into the decision-making process of the local agents were assessed. In the second study, the ability of the local agents in the proposed framework to learn policies aligned with the global decision of the coordinator agent was evaluated. In the last study, the framework was benchmarked against the learning-based multi-agent MPC for irrigation scheduling, which integrates MARL and MPC in a complementary fashion to address the daily irrigation problem. The following sections outline the specific details of these studies.

\subsection{Assessment of Communication and Augmentation Strategies}
The effectiveness of the communication and augmentation approaches was assessed by comparing the proposed SCMARL scheme with a Decentralized Multi-Agent Reinforcement Learning (DMARL) scheme. The DMARL approach also includes a coordinator agent that makes `yes/no' irrigation decisions based on global soil moisture, weather, and crop information. However, in the DMARL scheme, the coordinator's decisions are not shared with the local agents, and no augmentation is performed for the inputs of the local agents. Thus, in the DMARL framework, the coordinator's decision only serves as a switch to turn on or off the irrigation amounts prescribed by the local agents.

Under this study, the DMARL and SCMARL approaches were compared in two main ways. Firstly, the trained agents in the two approaches were evaluated using two main metrics: trajectories of average rewards over 9,800 episodes and 10 runs, and average training times over these 10 runs. Secondly, the DMARL and SCMARL approaches were employed to provide irrigation schedules for wheat crop in the study area over the 2022 growing season. The relevant weather information for the 2022 growing season used for this investigation can be found in Appendix~\ref{section:weather_data}. This investigation further compared the two approaches in terms of the total prescribed irrigation over the growing season and IWUE, defined as the ratio of the predicted yield of wheat to the total prescribed irrigation. The specific parameters and relations used during the calculation of the predicted yield of wheat are detailed in Appendix~\ref{sec:predicted_yield}. The subsequent sections will outline the specific details and settings under which this study was performed.

\subsubsection{Training and Initialization Settings of DMARL Agents}
In order to facilitate an even comparison between the two approaches, the parameters outlined in Section~\ref{sec:training_scmarl}  for training agents in the SCMARL approach were also applied to the agents in the DMARL approach. Additionally, the same method was used to initialize the weights of the actor and critic networks of the agents in the SCMARL AND DMARL frameworks.

\subsubsection{Simulation Settings of Season-long Investigation}\label{sec:season_long_init}
While the training of the agents in the SCMARL and DMARL approaches was performed in the fully observable MDP framework, the season-long investigation did not assume knowledge of soil moisture distribution in the MZs of the field. Instead, the investigation simulated the presence of a remote sensor in each MZ which was able to provide daily soil moisture observations that correspond to the average soil moisture content in the top 25 cm of the soil column. The soil moisture distribution, necessary for evaluating the agents in both frameworks, was estimated from daily soil moisture observations using the extended Kalman filtering (EKF) technique. The specific EKF design for each MZ can be found in Appendix~\ref{sec:ekf_design}.

The season-long investigation spanned from 15th May to 4th September, 2022. During the investigation, the spatial soil moisture content in the 3 MZs that make up the field was initialized with a guess of the soil moisture content in each zone. On the first evaluation day, these initial guesses, along with the daily weather and crop coefficient data, were used to evaluate the schedulers. The prescribed irrigation amounts were then applied to the actual field. The actual field conditions were represented using 3 well-calibrated 1D Richards equations, with added noise (mean of 0 and a standard deviation of 0.0007). On the second day, measurements corresponding to the average soil water content in the top 25 cm were obtained for each MZ from the actual field. To reflect sensor noise, additional noise from a normal distribution with a zero mean and a standard deviation of 0.0008 was added to these measurements. These noisy measurements were combined with the current soil moisture estimates in each MZ using the EKF approach to update the soil moisture estimates. Additionally, noise was added to the actual weather and crop coefficient data, and the noisy versions were used to evaluate the agents in the scheduling schemes while the actual weather and crop coefficient data were applied to the actual field to reflect the imperfections in weather and crop coefficient predictions. The updated soil moisture estimates, along with the weather and crop coefficient data for the second day of the evaluation period, were then used to prescribe the irrigation decision and irrigation amounts for the MZs of the field. This process was repeated daily until the end of the simulation period.

\subsection{Assessing Local Agents’ Alignment with Coordinator’s Decision}
This study assessed the ability of local agents in the SCMARL scheme to align their actions with the coordinator’s decisions. It examines how frequently the trained local agents prescribe actions that conform to the coordinator's decisions during a predefined set of evaluations of the SCMARL framework.

\subsection{Comparison between SCMARL and learning-based multi-agent MPC}
In this study, the SCMARL and learning-based multi-agent MPC approaches were employed to provide irrigation schedules for the crop in the study area over the 2022 growing season. The two approaches were compared in terms of total prescribed irrigation and IWUE. The subsequent sections provide a detailed description of the learning-based multi-agent MPC approach and the specific settings under which the investigation was performed.
\subsubsection{Learning-based multi-agent MPC: Description and Evaluation Settings}
\begin{figure}[H]
			\centering
			\includegraphics[width=0.60\columnwidth]{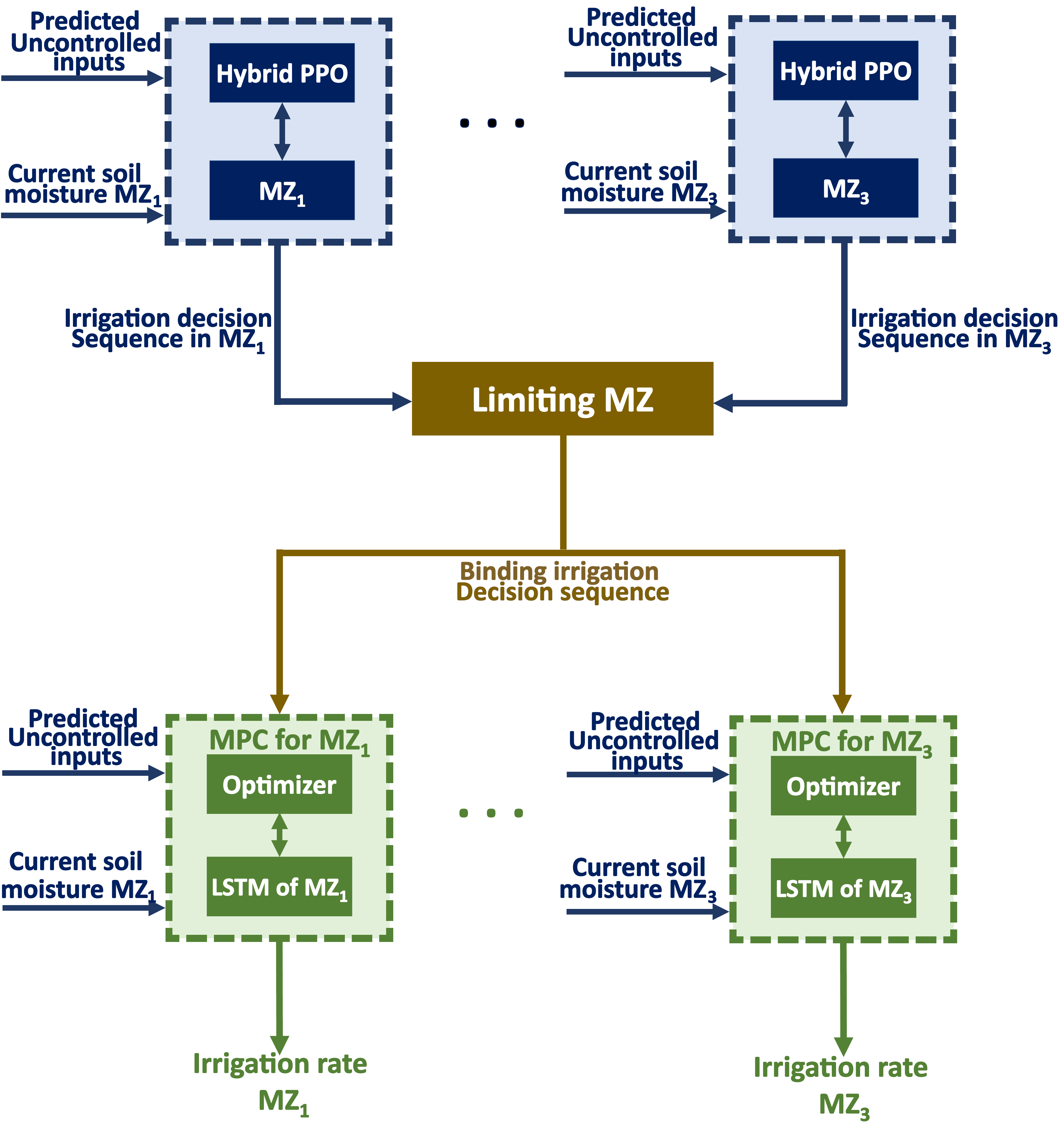}
	\caption{A schematic representation of the daily evaluation of the learning-based multi-agent MPC approach during the season-long investigation.}
	\label{fig:lb_ma_mpc_schem_diag}
\end{figure}

The learning-based multi-agent MPC approach, abbreviated as LB-MA-MPC, is depicted in Figure~\ref{fig:lb_ma_mpc_schem_diag}.  
In this approach, hybrid PPO agents were first trained in a decentralized manner for the various MZs that make up the field. For a particular MZ, the agent determines the irrigation decision and the corresponding irrigation amount that maintain its root zone soil moisture content within a target range while minimizing the fixed and variables irrigation costs.  Each agent considered the spatial soil moisture distribution of its respective MZ together with crop coefficient and weather data (reference evapotranspiration and rain) as input to its policy, and it interacted with a calibrated Richards equation during its training.

Due to the decentralized nature of the agents, discrepancies usually exist between the irrigation decisions calculated for the various MZs, which is inconsistent with the requirements of the daily irrigation scheduling problem. To handle this discrepancy, a heuristic approach is employed to determine a uniform irrigation decision that is applied to all MZs. In finding the uniform irrigation decision, each agent is initially evaluated over a predetermined prediction horizon using the soil moisture distribution on the first day of the horizon, as well as the predicted weather and crop information over the horizon. Note that the term predicted uncontrolled inputs is used to represent the crop data and weather predictions over the horizon in Figure~\ref{fig:lb_ma_mpc_schem_diag}. During this evaluation, each hybrid agent interacts with its respective environment, yielding irrigation decisions and corresponding irrigation amounts over the prediction horizon. The heuristic approach uses the irrigation decision sequences calculated by the various hybrid PPO agents to provide a uniform irrigation decision sequence (also referred to as a binding irrigation decision sequence in Figure~\ref{fig:lb_ma_mpc_schem_diag}) for all MZs. To determine the irrigation amounts consistent with this uniform irrigation decision sequence, parallel MPCs with continuous controls are solved over the considered horizon. In the solution of these MPCs, the uniform irrigation decision sequence is enforced, yielding the irrigation amounts for the various MZs. Interested readers may refer to~\cite{agyeman2024learning} for a detailed description of the LB-MA-MPC approach.

\subsubsection{Hybrid PPO Agent Training in Learning-based multi-agent MPC}
The weights used in the design of the reward functions for the agents in the LB-MA-MPC approach were adopted from those used in the SCMARL approach. Furthermore, the same hyperparameters used during the training of the agents in the SCMARL approach were applied to the training of the decentralized hybrid PPO agents. Additionally, a consistent neural network weight initialization technique was employed during the training of the agents in both scheduling approaches.

\subsubsection{Evaluation Setting of the SCMARL}
\begin{figure}[H]
			\centering
			\includegraphics[width=0.60\columnwidth]{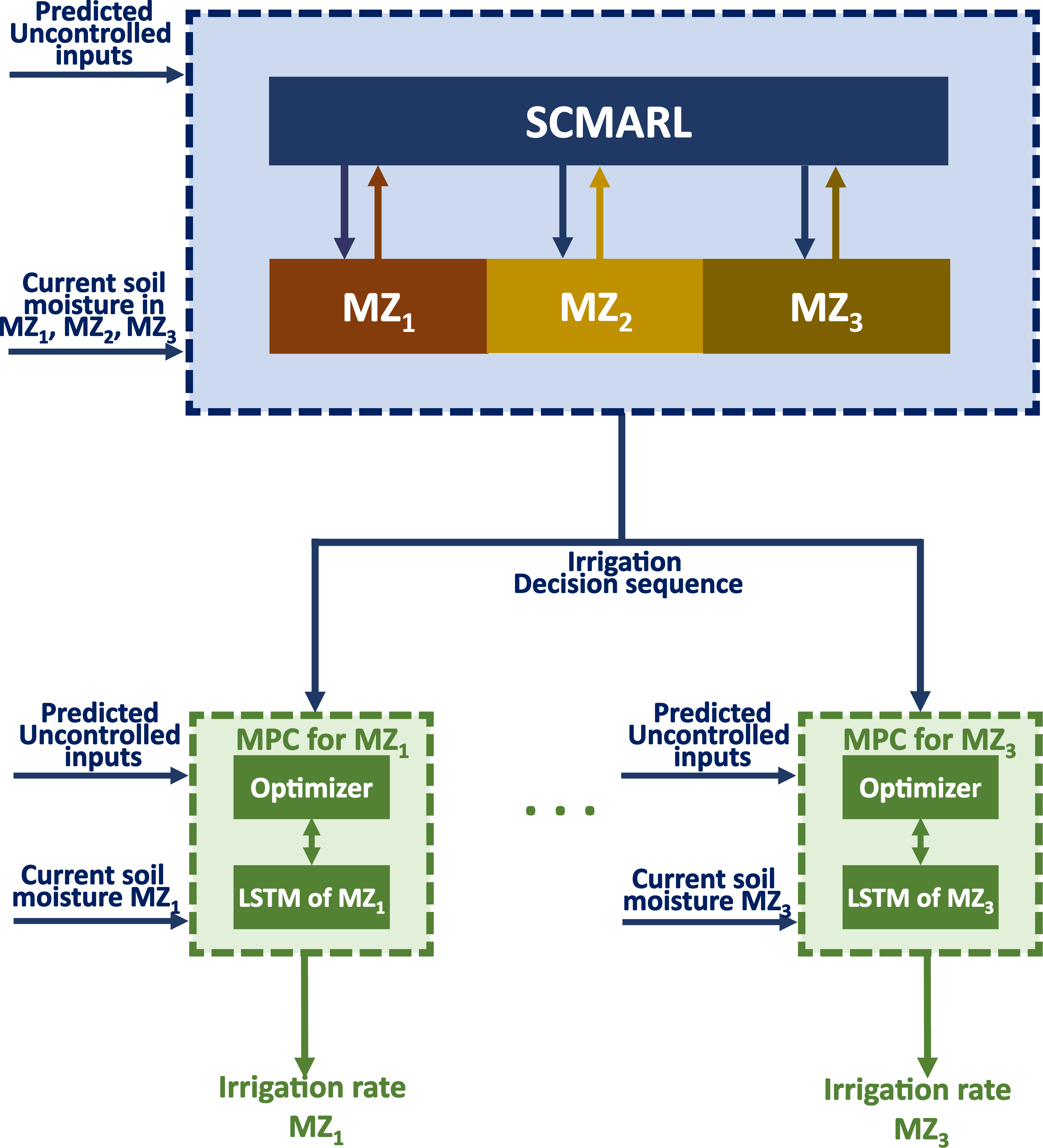}
	\caption{A schematic representation of the daily evaluation of the SCMARL framework  during the season-long investigation.}
	\label{fig:scmarl_schem_diag}
\end{figure}

To facilitate an even comparison with the LB-MA-MPC approach, the evaluation of the agents in the SCMARL for a particular day is modified. Similar to the LB-MA-MPC approach, the SCMARL approach is evaluated over a prediction horizon. A schematic representation of this daily evaluation is depicted Figure~\ref{fig:scmarl_schem_diag}. This evaluation uses the soil moisture distributions in the various MZs on the first day of the horizon, as well as crop data and weather predictions over the horizon. The term predicted uncontrolled inputs is used to represent the crop data and weather predictions over the horizon in Figure~\ref{fig:scmarl_schem_diag}. During the evaluation, the agents in the SCMARL interact with their respective environments, yielding the irrigation decisions and corresponding irrigation amounts for the MZs over the prediction horizon. 

The parallel MPCs solved in the LB-MA-MPC approach are employed under the SCMARL scheme to determine the final irrigation amounts applied to the field. However, in the solution of the MPCs under this approach, the binding irrigation decision sequence used in the LB-MA-MPC approach is replaced with the irrigation decision sequence obtained from the evaluation of the agents in the SCMARL approach.

\subsubsection{Simulation Settings of Season-Long Investigation}
Similar to the season-long investigation in  Section~\ref{sec:season_long_init}, exact knowledge of soil moisture distribution in the MZs of the field was not assumed. Instead, the presence of a remote sensor for each MZ, providing daily soil moisture observations corresponding to the average soil moisture content in the top 25 cm of the soil column in each zone, was simulated. The soil moisture distribution required for evaluating the agents in both approaches was estimated from these daily observations using the EKF. In this investigation, the design of the EKF for a particular MZ can be found in Appendix~\ref{sec:ekf_design}.

The investigation period spanned from May 15th to September 4th, 2022. A prediction horizon of 14 days was employed for executing the agents and solving the MPCs in both frameworks. Additionally, a control horizon of 14 days was used during the solution of the MPCs, which utilized Long-Short Term Memory (LSTM) networks to represent the soil moisture dynamics for each MZ. For a particular MZ, the formulation of the MPC that was solved to obtain the irrigation amounts can be found in Appendix~\ref{sec:mpc_design}. Interested readers may refer to~\cite{agyeman2024learning} for details regarding the design and training of the LSTM networks for the various MZs.

On the first day of the evaluation period, the agents in the two scheduling schemes were evaluated using the initial guess of the soil moisture distributions for the various MZs, along with the 14-day weather and crop information predictions. The uniform irrigation decision sequence, determined by evaluating the two scheduling schemes, was used in the MPC to find the irrigation amounts for the various MZs. Although the actual weather information for the growing season was known during this simulation, uncertainty was incorporated into the weather forecasts used in evaluating the agents and in the MPC. This uncertainty was modeled as a normal distribution with a mean of 0 and a specified standard deviation. As the prediction horizon extended further into the future, the standard deviation values were gradually increased to reflect the increasing uncertainty associated with longer-term weather predictions.

The irrigation amounts obtained from solving the MPC were implemented in a receding horizon fashion, where the first irrigation amount was applied to the actual field and the rest were discarded. The actual field conditions were represented using three well-calibrated 1D Richards equations, with added noise (mean of 0 and a standard deviation of 0.0007) to account for model uncertainties. On the second day, measurements corresponding to the average soil water content in the top 25 cm were obtained for each MZ from the actual field. To reflect sensor noise, additional noise from a normal distribution with a zero mean and a standard deviation of 0.0008 was added to these measurements. These noisy measurements were combined with the current soil moisture estimates in each MZ using the EKF to update the soil moisture estimates. The updated soil moisture estimates, along with the forecasted weather and crop coefficient data, were then used to prescribe the irrigation decisions and irrigation amounts for the MZs. This process was repeated daily until the end of the investigation period.

\section{Results and Discussion}
In this section, the results of various studies employed to assess the performance of the proposed SCMARL framework are presented and discussed in detail. The section begins with a presentation and discussion of the average score trajectories obtained during the training of the agents in the proposed framework. Next, the results of studies assessing the benefits of the communication and augmentation strategies are discussed. This is followed by a discussion of the study investigating the ability of the local agents to align their prescriptions with the coordinator agent's decision. Finally, the irrigation schedules and corresponding root zone soil moisture trajectories obtained under the proposed and the learning-based multi-agent MPC scheduling approaches during the season-long investigation are presented and discussed in detail.

\subsection{Training Performance of SCMARL Agents}
\begin{figure}[H]
	\centering
	\includegraphics[width=0.6\columnwidth]{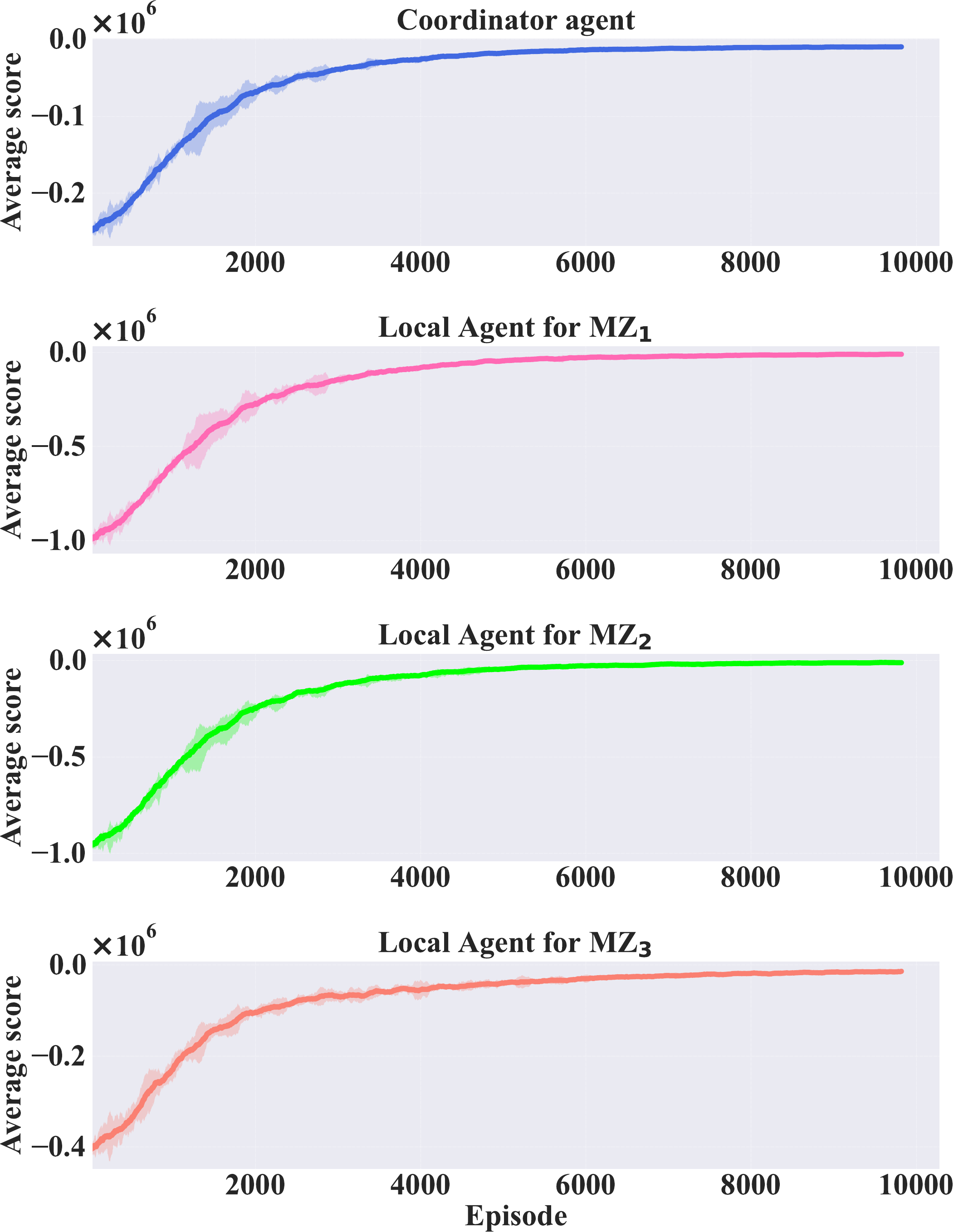}
	\caption{Average reward trajectories of the agents over 9800 episodes.}
	\label{fig:rew_traj}
\end{figure}

Figure~\ref{fig:rew_traj} illustrates the average score obtained from running each of the four agents through 10 repetitions across a span of 9800 episodes. During the initial training phase, extending up to approximately the 6000th episode, the average score trajectories for all agents generally display an upward trend. After the 6000th episode, the average score trajectory remains  fairly constant for all the agents. 
The steady increase in average scores during the initial phase is indicative of effective learning. Furthermore, the stabilization of the score trajectories after the 6000th episode suggests that the agents have reached a point where they consistently apply near-optimal/optimal actions for irrigation scheduling. Overall, Figure~\ref{fig:rew_traj} highlights the effectiveness of the SCMARL framework in enabling the agents to converge to efficient irrigation strategies over time.

\subsection{Performance of Communication and Augmentation Strategies}
\begin{figure}[H]
	\subfloat[Average reward over 9800 episodes]{
		\begin{minipage}[c]{0.50\columnwidth}
			\centering
			\includegraphics[width=0.95\columnwidth]{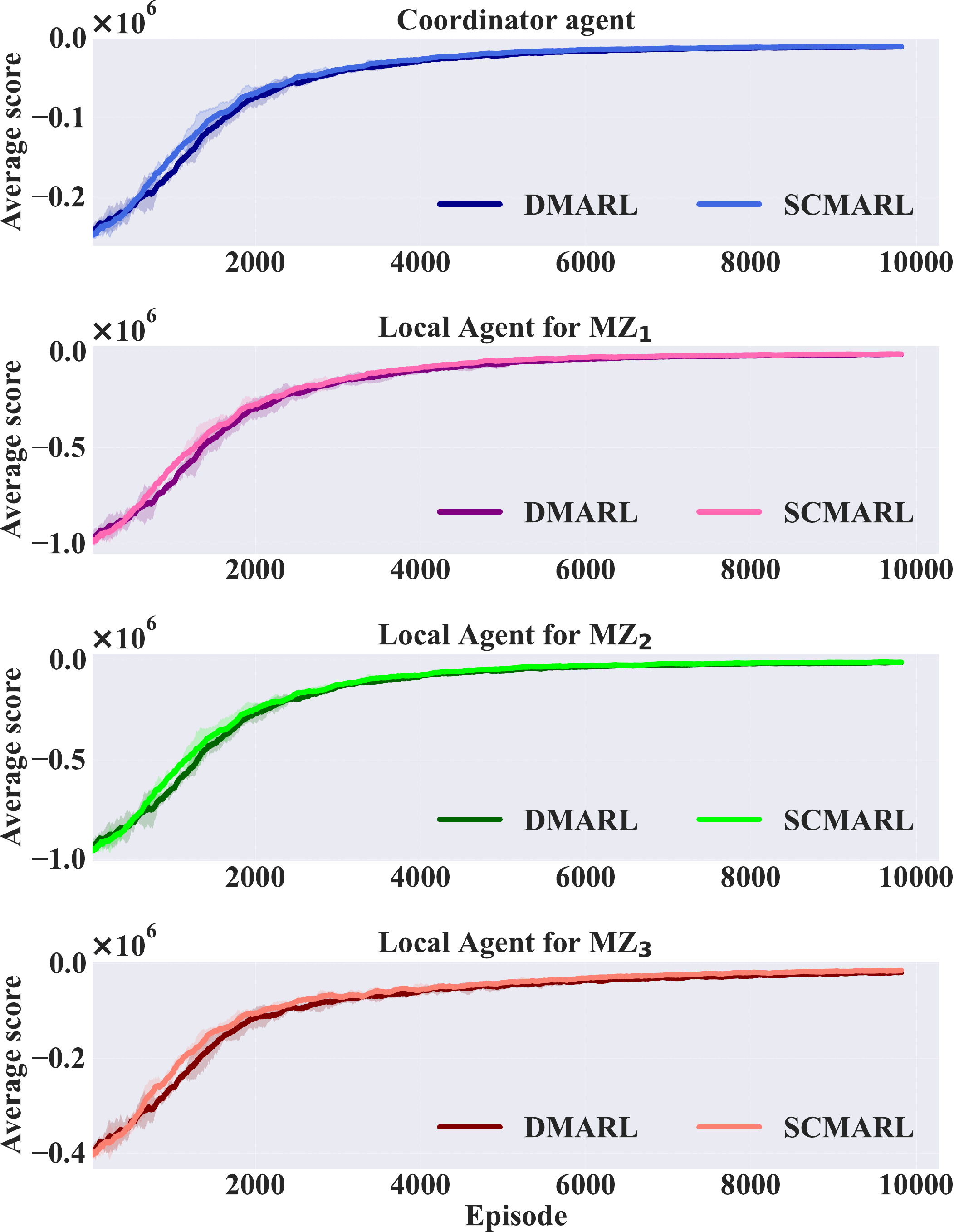}
	\end{minipage}}
	\hfill
	\subfloat[Average reward from 9000 to 9800 episodes.]{
		\begin{minipage}[c]{0.50\columnwidth}
			\centering
			\includegraphics[width=0.95\columnwidth]{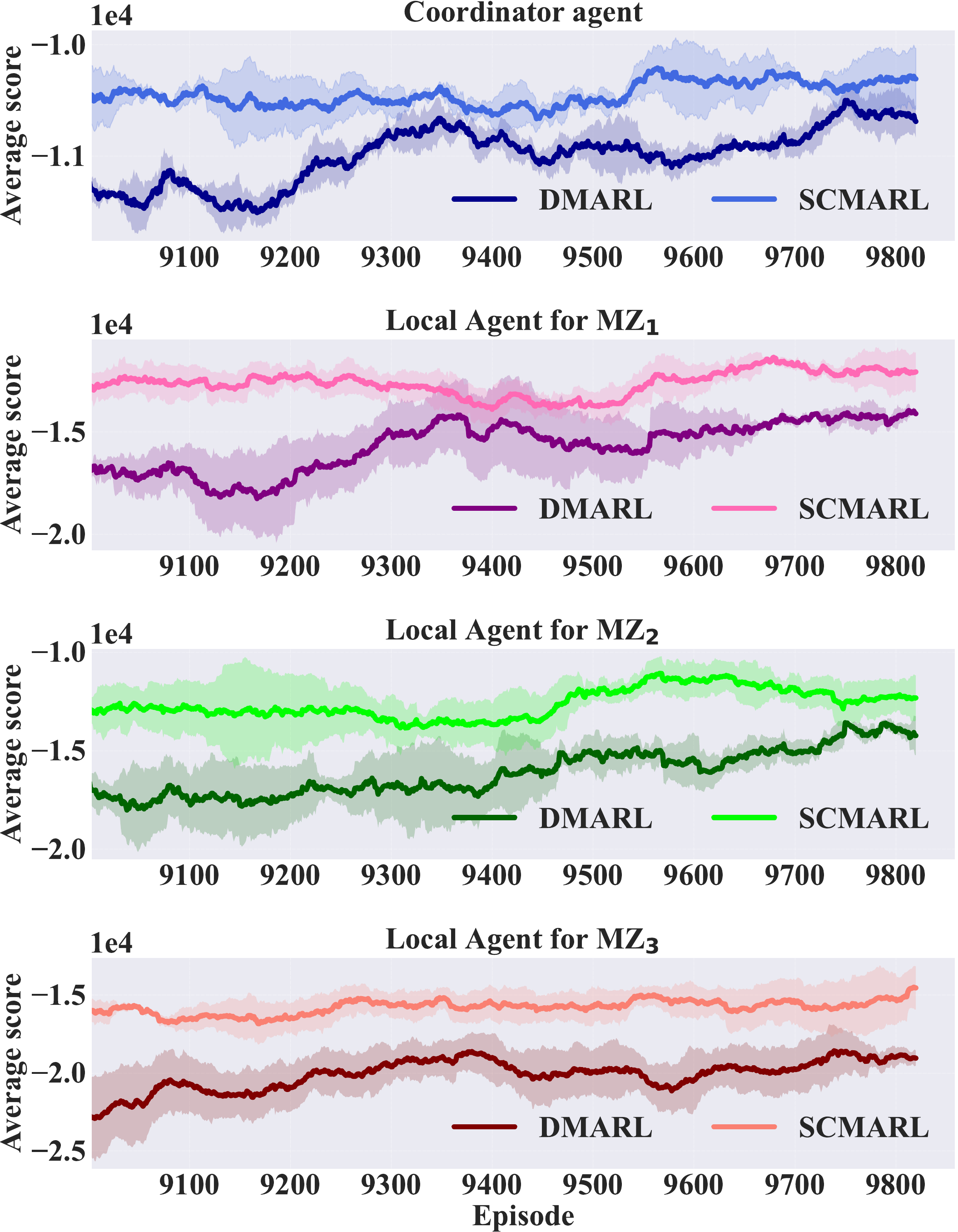}
	\end{minipage}}
	\caption{Comparison between the SCMARL and the DMARL approaches}
	\label{fig:scmarl_vs_dmarl_rew_traj}
\end{figure}

Figure~\ref{fig:scmarl_vs_dmarl_rew_traj} compares the average score trajectories of agents in the SCMARL approach with those in the DMARL approach. The plots in Figure~\ref{fig:scmarl_vs_dmarl_rew_traj}(a) displays the average score for the agents across 9800 episodes over 10 runs. In the initial training phase, extending up to the 6000th episode, the agents in the SCMARL approach show a more consistent and higher average score trajectory compared to the DMARL approach. This difference  is particularly more evident in the performance of the local agents.
To provide a detailed view of the convergence phase, Figure~\ref{fig:scmarl_vs_dmarl_rew_traj}(b) focuses on the last 800 episodes (from 9000 to 98000). This close-up view highlights the SCMARL agents superior performance, with average scores stabilizing at higher values compared to the agents in the DMARL approach.

\begin{figure}[H]
	\centering
	\includegraphics[width=0.9\textwidth]{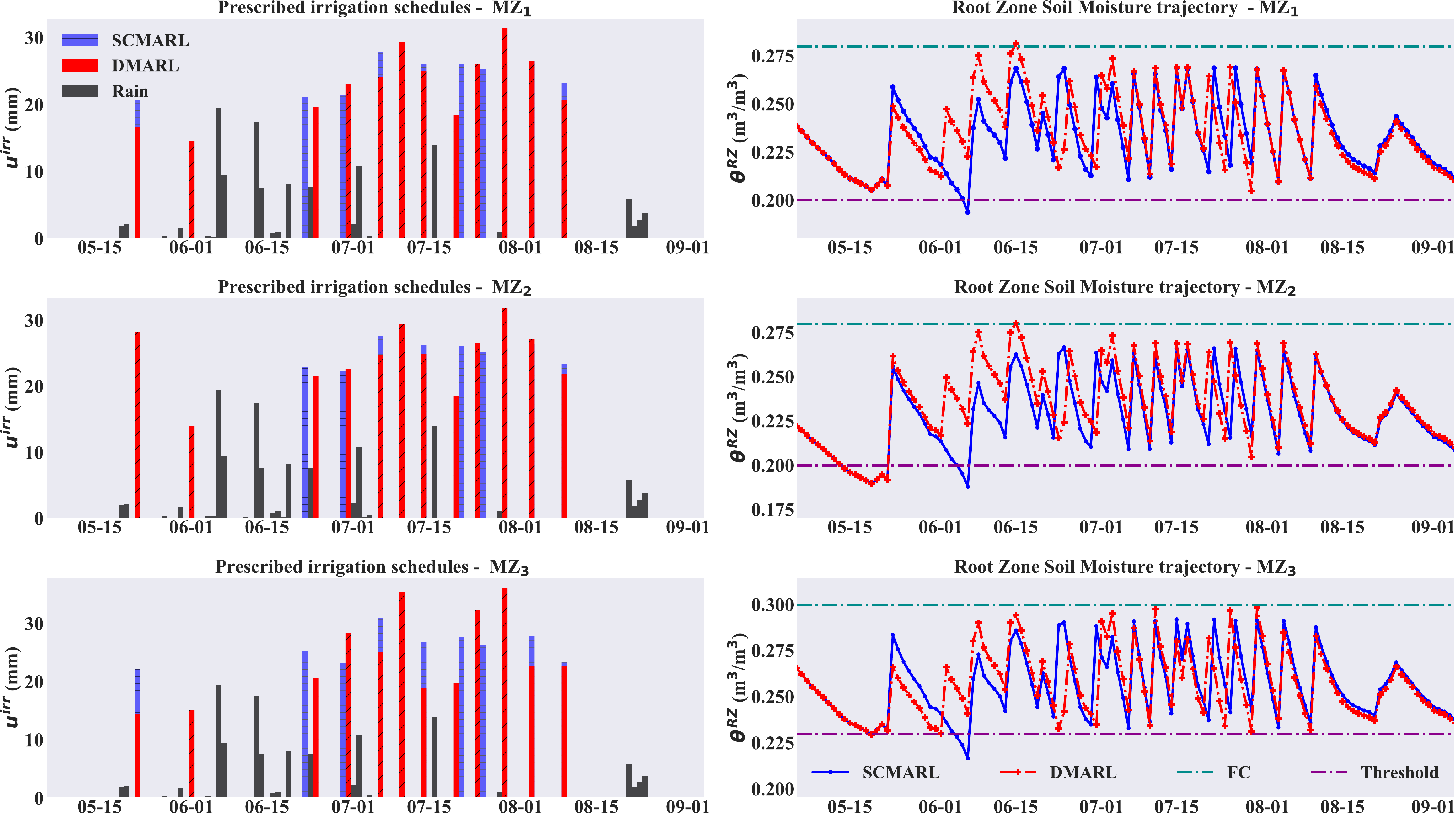}
	\caption{Prescribed irrigation schedules and the trajectories of root zone soil moisture content under the SCMARL and DMARL schemes.}
	\label{fig:schd_res_scmarl_vs_dmarl}
\end{figure}

Figure~\ref{fig:schd_res_scmarl_vs_dmarl} shows the root zone soil moisture content in the 3 MZs of the field under the schedules prescribed by the SCMARL and DMARL scheduling schemes during the season-long investigation. It is evident from Figure~\ref{fig:schd_res_scmarl_vs_dmarl} the two schemes are able to provide schedules that maintain the root zone soil moisture content within the target soil moisture range for all the MZs that make up the field. The total prescribed irrigation and the IWUE obtained from the two scheduling schemes are shown in Table~\ref{tbl:comp_scmarl_vs_dmarl}. From Table~\ref{tbl:comp_scmarl_vs_dmarl}, the SCMARL approach prescribed a lower total irrigation, compared to the DMARL approach. Additionally, the SCMARL approach achieved a higher IWUE. Using the DMARL approach as the benchmark, the SCMARL approach achieved a 1.9\% savings in water used for irrigation while enhancing the IWUE by 1.9\%. These results further highlight the superior performance of the SCMARL approach compared to the DMARL approach.

The superior performance of the SCMARL approach compared to the DMARL approach is attributable to the communication of the coordinator's decision to the local agents and its subsequent inclusion in the local agents' policy inputs. This method effectively reduces uncertainties in the environments of the local agents, leading to more stable learning conditions. These results align with findings in~\cite{zhou2021hierarchical}, where a hierarchical structure with directed communication was shown to handle non-stationarity effectively in MARL applications within real-time strategy games.

While the SCMARL approach performed better than the DMARL approach, the performance of the agents in the DMARL approach remains acceptable. The decentralized training approach in DMARL is similar to independent learning approaches applied in MARL studies such as~\cite{tampuu2017multiagent} and~\cite{de2020independent}. The ability of the DMARL approach to achieve efficient learning can be attributed to two main reasons. Firstly, as reported in~\cite{nekoei2023dealing}, a commonly used scheme for decentralized deep MARL is to approximate what is termed independent iterative best response, where agents independently and concurrently try to find the best response strategy with respect to other agents’ policies. In the DMARL approach, this translates to local agents trying to find the best response to the coordinator agent's policy. Secondly, the PPO algorithm used in the DMARL approach mitigates some forms of environmental non-stationarity, as shown in~\cite{de2020independent}, through the policy clipping technique. 

Table~\ref{tbl:scmarl_vs_dmarl_time} shows that the training time for the DMARL approach is shorter compared to the SCMARL approach. This difference is due to the additional computational overhead introduced by the communication of the coordinator's decision to the local agents in the SCMARL approach. Additionally, the augmentation step in SCMARL results in more complex policies for the local agents, requiring more time to update and further extending the training duration compared to the simpler policies in DMARL. Consequently, in instances where faster training time is crucial, the DMARL approach may be preferred for the irrigation scheduling problem. However, as reported in~\cite{nekoei2023dealing}, the independent iterative best response approach used in DMARL can fail to converge to agent-by-agent optimal solutions. Further studies are necessary to identify the specific settings under which the DMARL approach can reliably achieve near-optimal or optimal policies, particularly in the context of daily irrigation scheduling.

\subsection{Local Agents' Policy Alignment with Coordinator's Decision}
Table~\ref{tbl:autonomy_test} confirms that the communication and augmentation strategies, coupled with the switch-like implementation of the coordinator's decision during training, effectively encourage local agents to develop policies that align with the global irrigation decision (coordinator's action). In 98\% of evaluations, local agents in the SCMARL approach prescribed actions that matched the coordinator's decision, eliminating the need to explicitly enforce the coordinator's decision during execution. However, the local agents failed to align their actions with the global strategy in 33 out of 2000 runs. Extending training episodes could reduce this failure rate. Additionally, employing a predefined threshold irrigation amount can help mitigate the impact of local agent actions that do not align with the global strategy, as it may not be feasible to eliminate such failures entirely.

\begin{table}[H]
	\caption{Comparison between SCMARL and DMARL approaches.}
	\centering
	\begin{tabular}{cccc}
		\toprule
		\textbf{Metric} & \textbf{SCMARL}&\textbf{DMARL} \\
		\midrule
		Total irrigation (m) & 0.785  $\bm{\left[{\downarrow}\textbf{1.9}\%\right]}$ & 0.800\\ 
		IWUE (kg/m$^3$) & 1.118 $\bm{\left[{\uparrow}\textbf{1.9}\%\right]}$& 1.097 \\
		\bottomrule
	\end{tabular} \label{tbl:comp_scmarl_vs_dmarl}
\end{table}

\begin{table}[H]
	\caption{Average time required to train agents in the  SCMARL and DMARL approaches.}
	\centering
	\begin{tabular}{cccc}
		\toprule
		\textbf{Metric} & \textbf{SCMARL}&\textbf{DMARL} \\ 
		\midrule
		Average Time (hours) & 9.2 & 8.8\\ 
		\bottomrule
	\end{tabular} \label{tbl:scmarl_vs_dmarl_time}
\end{table}

\begin{table}[H]
	\caption{Assessing the autonomy of the local agents resulting from the communication and augmentation approaches.}
	\centering
	\begin{tabular}{cccc}
		\toprule
		\textbf{Number of Evaluations} & \textbf{Number of Successes}&\textbf{Number of Failures} \\ 
		\midrule
		2000 & 1967 (\textbf{98\%})  & 33 (\textbf{2\%}) \\ 
		\bottomrule
	\end{tabular} \label{tbl:autonomy_test}
\end{table}

\subsection{Performance and Utility of the SCMARL Framework}

\begin{figure}[H]
	\centering
	\includegraphics[width=0.9\textwidth]{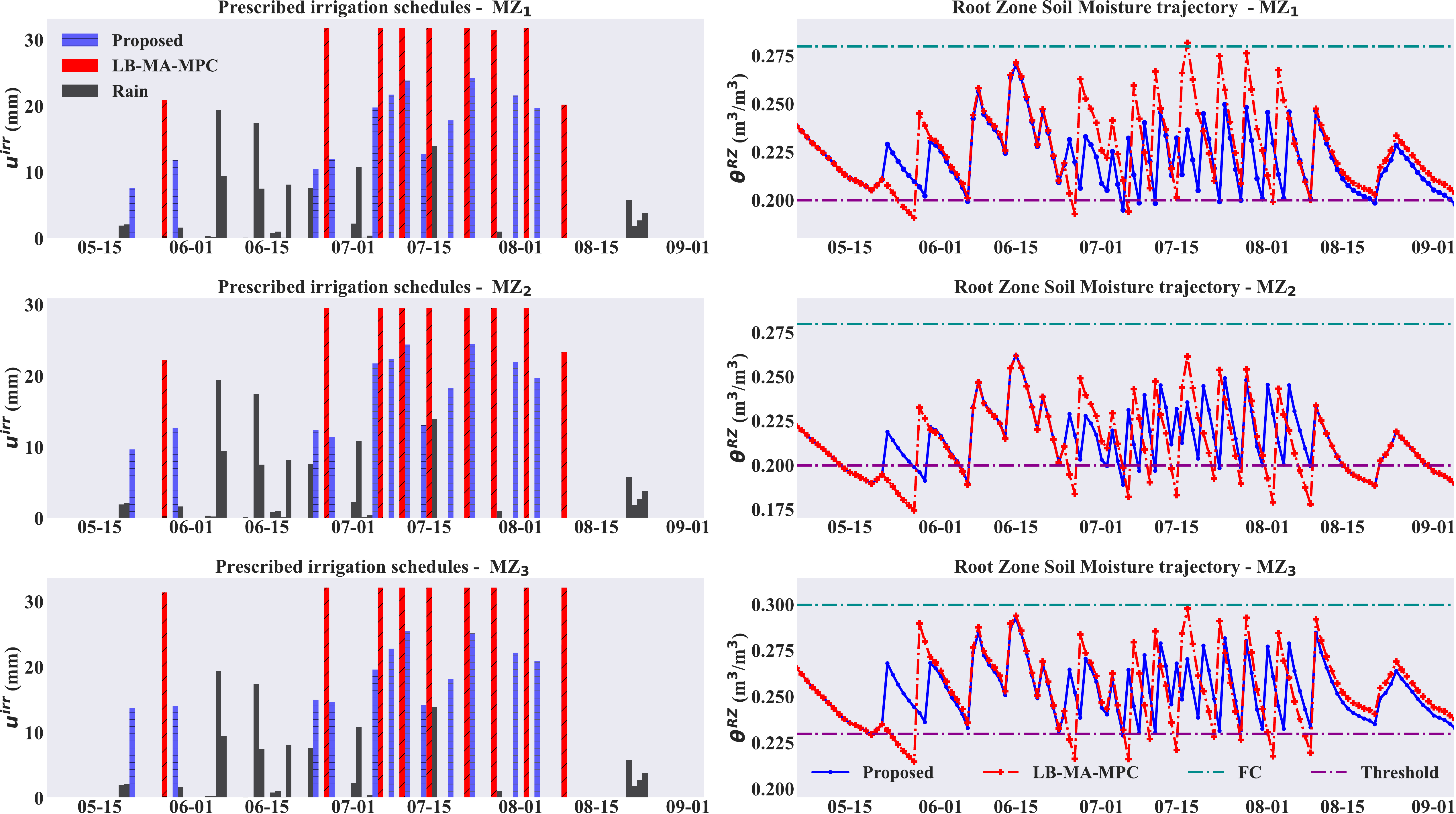}
	\caption{Prescribed irrigation schedules and the trajectories of root zone soil moisture content under SCMARL and the learning-based multi-agent MPC approaches.}
	\label{fig:schd_res_scmarl_vs_lb_mpc}
\end{figure}

\begin{table}[H]
	\caption{Comparison between SCMARL and the learning-based multi-agent MPC scheduling approaches.}
	\centering
	\small
	\begin{tabular}{cccc}
		\toprule
		\textbf{Metric} & \textbf{SCMARL}&\textbf{LB-MA-MPC} \\ 
		\midrule
		Total irrigation (m) & 0.774  $\bm{\left[{\downarrow}\textbf{4.0}\%\right]}$ & 0.806\\

		IWUE (kg/m$^3$) & 1.134 $\bm{\left[{\uparrow}\textbf{6.3}\%\right]}$& 1.067 \\ 
		\bottomrule
	\end{tabular} \label{tbl:comp_scmarl_vs_lb_mpc}
\end{table}

Figure~\ref{fig:schd_res_scmarl_vs_lb_mpc} illustrates the trajectories of root zone soil moisture content in the three MZs under the schedules recommended by the trained agents in the SCMARL and the LB-MA-MPC scheduling approaches. This figure demonstrates that both scheduling schemes are able to maintain the root zone soil moisture content within the target range, with occasional violations of the bounds for each MZ of the field.

Table~\ref{tbl:comp_scmarl_vs_lb_mpc} presents a quantitative comparative analysis between the proposed SCMARL and the LB-MA-MPC scheduling approaches. This table reveals that the proposed SCMARL approach prescribed a lower total irrigation amount compared to the LB-MA-MPC approach, resulting in a 4.0\% reduction in total irrigation amount. Additionally, in terms of IWUE, the proposed SCMARL approach achieved a higher IWUE, specifically a 6.3\% increase compared to the LB-MA-MPC approach.

Based on the evaluation of the proposed SCMARL framework and the LB-MA-MPC approach, the results reported in Table~\ref{tbl:comp_scmarl_vs_lb_mpc} highlight the significant impact of the heuristic approach used in the LB-MA-MPC method to unify the daily irrigation decisions that are obtained from the decentralized agents. This suboptimal approach leads the LB-MA-MPC to recommend higher daily irrigation amounts less frequently, causing violations of the target soil moisture range and negatively affecting crop yield, as reflected in the lower IWUE.
In contrast, the proposed SCMARL approach, which considers a comprehensive view of the entire field, enhances the optimality of the irrigation decision. Using optimal irrigation decisions to determine irrigation amounts with decentralized MPCs results in irrigation schedules that improve water conservation and crop yields.

Compared to the learning-based multi-agent MPC approach, where the decentralized MPC must be solved to determine the final irrigation amounts, the inherent design of the SCMARL framework shows that the decentralized MPC is not a major requirement for its implementation. The irrigation amounts computed by the local agents in the SCMARL framework align with the global irrigation decision, reducing the necessity for a decentralized MPC step.
However, comparing  Tables~\ref{tbl:comp_scmarl_vs_dmarl} and~\ref{tbl:comp_scmarl_vs_lb_mpc} reveals that combining the SCMARL approach with the decentralized MPC enhances the overall ability of the combined framework to achieve additional water savings and improve IWUE. By employing the decentralized MPC, the proposed approach achieved an additional 1.4\% water savings compared to the case without the decentralized MPC step. Furthermore, employing the decentralized MPC in the proposed framework enhanced the IWUE by 1.4\%. Note that this comparison is possible since the same actual field representation, estimator design, weather data, and noise statistics were adopted in the two season-long simulations that produced the results summarized in Tables~\ref{tbl:comp_scmarl_vs_dmarl} and~\ref{tbl:comp_scmarl_vs_lb_mpc}.

The above observation highlights that RL and MPC can be employed in a complementary manner, corroborating existing studies that have advocated and demonstrated the benefits of employing both approaches together~\cite{ernst2008reinforcement, negenborn2005learning}. Notably, the learning-based multi-agent MPC approach is the first known work in the area of irrigation scheduling that seeks to employ RL and MPC in a complementary manner. The results of this work demonstrate that combining the SCMARL approach with MPC results in better performance compared to the learning-based multi-agent MPC approach. This improvement was found to be  primarily due to the field-wide optimality of daily irrigation decision calculations in the SCMARL approach. Additionally, the local agents in the SCMARL approach provide irrigation amounts that align with the irrigation decision sequence, serving as better initial guesses for the MPC step compared to the decentralized agents in the learning-based approach. This combination can be particularly helpful since using the SCMARL framework alone for irrigation scheduling presents challenges such as computational overhead and complexity in policy design. It also lacks the precision in irrigation amounts that MPC provides. Additionally, redesigning the reward structure in the SCMARL framework would require retraining the entire system, which can be time-consuming compared to solving an MPC with a redesigned cost function. On the other hand, using MPC alone for irrigation scheduling, as employed in~\cite{agyeman2023lstm}, can be computationally demanding, especially when solving for both discrete and continuous variables simultaneously. Combining SCMARL with MPC leverages the strengths of both approaches: the learning capabilities and adaptability of RL to handle dynamic environments and the precision and optimization ability of MPC to fine-tune irrigation amounts, leading to a more computationally efficient and effective irrigation scheduling solution.

\section{Conclusion}
This paper proposed a two-tier semi-centralized MARL framework to address daily irrigation scheduling in spatially variable fields characterized by irrigation MZs. At the top level is a coordinator agent responsible for determining daily `yes/no' irrigation decisions based on field-wide soil moisture, weather, and crop information. The second level consists of local agents assigned to specific MZs, determining daily irrigation amounts based on local conditions and the coordinator's decision. The framework employs communication and state augmentation strategies to handle non-stationarity, leading to stable learning environments for the local agents.

A comparison of the proposed approach with a fully decentralized framework indicated that the communication and augmentation strategies reduced uncertainty in the environments of the local agents, resulting in stable and improved learning. Applying the proposed approach to a large-scale field demonstrated its capability to achieve substantial water savings and improved irrigation water use efficiency compared to a learning-based MPC approach, which employs MARL and MPC in a complementary manner to address the daily irrigation scheduling problem. Furthermore, while the proposed approach permits the use of MARL alone to address the daily irrigation scheduling problem, the results indicated that employing the SCMARL framework with MPC in a complementary manner enhances the performance of the proposed framework.

While the proposed approach shows promise, several modifications could enhance its effectiveness. Although the scalability of the proposed approach to a field with several MZs is superior to that of a single, centralized agent, the coordinator agent, which is centralized, needs modifications when the proposed approach is applied to fields with a medium to high number of MZs. In the present approach, the entire soil moisture distribution of the soil column under consideration is included as one of the inputs to the policies of all the agents in the framework. While this approach appears acceptable for the local agents, it results in a situation where the input space of the coordinator agent becomes large when a field with a medium to high number of MZs is considered. This increases the complexity of the coordinator agent's policy and further slows the training time of the agents in the framework. One possible modification to manage the input space of the coordinator agent in such an instance is to replace the entire soil moisture distribution in the inputs of the policies with the rootzone soil moisture content in the various MZs. Since the rootzone soil moisture is a scalar value computed based on the entire soil moisture distribution, employing it in place of the entire soil moisture distribution can significantly reduce the input space of the agents, especially the coordinator agent. This approach, however, can impact the overall performance of the proposed approach, as using a scalar value to represent the entire soil moisture distribution can result in the loss of information that could help the agents make optimal decisions.

Another potential modification is redesigning the agents in the framework to incorporate future weather and crop predictions as inputs. In the present framework, the agents solely employ current weather and crop information in their decision-making process.

Lastly, investigating the proposed framework within a partially observable MDP (POMDP) framework is another modification worth considering, as such a setting provides a more practical framework for agricultural fields where observations/soil moisture measurements of the MZs are readily available, instead of the entire soil moisture distribution in the various MZs. In this regard, the EKF can be employed, together with well-calibrated Richards equations for the various MZs, to estimate the soil moisture distribution based on the soil moisture observations obtained from the MZs. This estimated soil moisture can then be used as the belief state in a POMDP framework.

\section{Acknowledgment} 

Financial support from Natural Sciences and Engineering Research Council of Canada (NSERC) is gratefully acknowledge.


\appendix
\numberwithin{equation}{section}
\renewcommand{\thesection}{\Alph{section}.\arabic{section}}
\renewcommand{\thetable}{\Alph{section}.\arabic{table}}
\renewcommand\thefigure{\thesection.\arabic{figure}}
\setcounter{section}{0}
\setcounter{table}{0}
\setcounter{figure}{0}

\begin{appendices}
\section{MPC Formulation}\label{sec:mpc_design}
Using the irrigation decision sequence $\bm{c}$ obtained from evaluating the SCMARL agents or the hybrid PPO agents in the learning-based multi-agent MPC approach, the MPC for prediction horizon $N_p$ and day $d$ is formulated for a particular MZ as follows:
\begin{subequations}
	\begin{alignat}{2}
		\min_{\bm{\bar{\epsilon}},~\bm{\underline{\epsilon}},~\bm{u}^{\text{irrig}}} ~ &\sum_{k=d+1}^{d+N_p}\left[\bar{Q}\bar{\epsilon}^2_k + \underline{Q}\underline{\epsilon}^2_k \right] + \sum_{k=d}^{d+N_p-1}R_uu_k^{\text{irrig}}&& \label{eq:obj} \\
		\notag
		&\textrm{s.t. }&&\\ 
		&\theta^{\text{RZ}}_{k+1} = \mathcal{F}_{\text{LSTM}}(\{{\gamma}\}_{k-4}^{k},\eta) &&k\in [d,d+N_p-1] \label{eq:cons1}\\
		&\theta^{\text{RZ}}_d = \mathcal{W}(\hat{y}(d)) \label{eq:cons2} \\
		&\underline{\nu} - \underline{\epsilon}_k\leq \theta^{\text{RZ}}_k \leq \bar{\nu} + \bar{\epsilon}_k, && k\in [d+1,d+N_p] \label{eq:cons3} \\
		&c_k\underline{u}^{\text{irrig}} \leq u_k^{\text{irrig}} \leq c_k\bar{u}^{\text{irrig}}, && k\in [d,d+N_p-1] \label{eq:cons4} \\
		&\underline{\epsilon}_k \geq 0, \quad \bar{\epsilon}_k\geq 0, && k\in [d+1,d+N_p] \label{eq:cons6} 
	\end{alignat}
\end{subequations}
where $k \in \mathbb{Z}^+$, $\bm{\bar{\epsilon}}\coloneqq [ \bar{\epsilon}_{d+1}, \bar{\epsilon}_{d+2},...,\bar{\epsilon}_{d+N_p}]$, $\bm{\underline{\epsilon}}\coloneqq [ \underline{\epsilon}_{d+1}, \underline{\epsilon}_{d+2},...,\underline{\epsilon}_{d+N_p}]$, $\bm{c}\coloneqq [ c_{d}, c_{d+1},...,c_{d+N_p-1}]$, $c_k \in \bm{c}$ $\bm{u}^{\text{irrig}}\coloneqq [ u_{d}^{\text{irrig}}, u_{d+1}^{\text{irrig}},...,u_{d+N_p-1}^{\text{irrig}}]$, and $\{\gamma\}_{k-4}^{k} \coloneqq [ \gamma_{k-4}, \gamma_{k-3},..,\gamma_{k}]$ where $\gamma\in[\text{K}_{\text{c}}, \text{ET}_0, u^{\text{irrig}}, \text{z}_{\text{r}}]$. 
$\underline{\epsilon}_k$ and $\bar{\epsilon}_k$ are slack variables that are introduced to relax $\underline{\nu}$ and $\bar{\nu}$. 
$\eta$ in Constraint~(\ref{eq:cons1}) represents the weights and bias terms of  the LSTM network. The initial $\theta^{\text{RZ}}$ is represented with Constraint~(\ref{eq:cons2}). Its value is calculated based on the estimated soil moisture distribution $\hat{y}$ on day $d$. $\mathcal{W}$ in  Constraint~(\ref{eq:cons2}), is a compact representation of the approach outlined in Section~\ref{sec:reward_design} that is used to calculate $\theta^{\text{RZ}}$ from $y$. When $c_k = 0 $, Constraint~(\ref{eq:cons4}) requires that  $ u_k^{\text{irrig}} = 0$.  Conversely, when $c_k = 1 $, Constraint~(\ref{eq:cons4}) specifies that $u_k^{\text{irrig}} \in [\underline{u}^{\text{irrig}}, \bar{u}^{\text{irrig}}] $.
 
 \section{Extended Kalman Filter Design}\label{sec:ekf_design}
For a particular MZ, the EKF is designed as follows:

\textbf{Initialization}

 The EKF is initialized with a guess of the state vector $x(0)$, its covariance matrix $P(0|0)=15.9\mathbb{I}_{21}$,  covariance of the process disturbance $Q=0.05\mathbb{I}_{21}$, and the covariance of the measurement noise $R=19.25$.  
 
 \textbf{Prediction Step}

 At time instant $k+1$, $x$ and $P$ are predicted as follows:
\begin{equation}
	\label{eq:prediction}
	\hat{x}(k+1|k)=\mathcal{F}(\hat{x}(k|k),u(k),\bm{\phi})
\end{equation}
\begin{equation}\label{eq:propagation}
	P(k+1|k) = A(k) P(k|k) A(k)^T + Q
\end{equation}
where $A(k)=\frac{\partial \mathcal{F}}{\partial x}\big|_{\hat{x}(k|k),~u(k)}$ 

\textbf{Update Step} 

Using the soil moisture observation $o(k+1)$ at time $k+1$, $x$ and $P$ are updated as follows:

\begin{equation}
	\begin{aligned}
		\hat{y}(k+1|k)=\mathcal{H}(\hat{x}(k+1|k),\bm{\phi})
	\end{aligned}
\end{equation}
\begin{equation}
	\begin{aligned}
		G(k+1) = P(k+1|k) C^T(k+1)[C(k+1)P(k+1|k)C^T(k+1) + R]^{-1}
	\end{aligned}
\end{equation}
\begin{equation}
	\begin{aligned}
		\hat{x}(k+1|k+1)=\hat{x}(k+1|k) + G(k+1)[o(k+1)-{\mathcal{M}}\hat{y}(k+1|k)]
	\end{aligned}
\end{equation}
\begin{equation}
	P(k+1|k+1)=[I-G(k+1)C(k+1)]P(k+1|k)
\end{equation}
\begin{equation}
	\begin{aligned}
		\hat{y}(k+1|k+1)=\mathcal{H}(\hat{x}(k+1|k+1),\bm{\phi})
	\end{aligned}
\end{equation}
where $\mathcal{M}$ serves as a selection matrix utilized to choose the soil moisture contents in $\hat{y}$ that collectively contribute to the soil moisture observation $o$ and $C(k+1)=\frac{\partial \mathcal{H}}{\partial x}\big|_{\hat{x}(k+1|k)}$.
\section{Predicted Yield Calculation}\label{sec:predicted_yield}
Crop yield is predicted according to the following equation~\cite{bennett2011crop}:
\begin{equation}
	\label{eq:yield_eqn}
	Y_a =Y_m\left[1- k_y + \left(k_y\times \frac{\text{ET}_{\text{c}}}{\text{ET}_{\text{m}}}\right)\right]
\end{equation}
where $Y_a$ is the predicted yield in (kg m$^{-2}$), $Y_m$ is the maximum potential yield in (kg m$^{-2}$),  $\text{ET}_{\text{c}}$ is seasonal crop evapotranspiration (mm), $\text{ET}_{\text{m}}$ is maximum seasonal crop evapotranspiration (mm), and $k_y$ is a crop-specific yield response factor (dimensionless). $\text{ET}_{\text{c}}$ is related to $\text{ET}_{\text{m}}$ as follows~\cite{feddes1982simulation}:
\begin{equation}
	\text{ET}_{\text{c}} = \mathcal{K}(\theta_{\text{v}})\text{ET}_{\text{m}}
\end{equation}
where $\mathcal{K}(\cdot)$ is the water stress factor,  which is a function of the volumetric moisture content $\theta_v$. $\mathcal{K}(\cdot)$ is defined as:
\begin{equation}
	\mathcal{K}(\theta_{\text{v}})=\begin{cases}
		0 & \text{$ \theta_{{\text{v}}_1} \leq \theta_{\text{v}}$}\\
		\frac{\theta_{\text{v}} - \theta_{{\text{v}}_1}}{\theta_{{\text{v}}_2}-\theta_{{\text{v}}_1}} &\text{$ \theta_{{\text{v}}_1} \leq \theta_{\text{v}} \leq \theta_{{\text{v}}_2} $}\\
		1& \text{$ \theta_{{\text{v}}_2} \leq \theta_{\text{v}}\leq\theta_{{\text{v}}_3} $}\\
		\frac{\theta_{\text{v}}-\theta_{{\text{v}}_w}}{\theta_{{\text{v}}_2}-\theta_{{\text{v}}_w}}&\text{$ \theta_{{\text{v}}_w}\leq \theta_{\text{v}}\leq \theta_{{\text{v}}_2} $}
	\end{cases}
\end{equation}
where $\theta_{{\text{v}}_1}$ is the volumetric moisture at the anaerobic point, $\theta_{{\text{v}}_2}$ and $\theta_{{\text{v}}_3}$ are the volumetric moisture content values between which optimal water uptake exists, and $\theta_{{\text{v}}_w}$ is  volumetric moisture content at the wilting point $\theta_{\text{wp}}$. $\theta_{{\text{v}}_3} = \bar{\nu}$, $\theta_{{\text{v}}_2} = \underline{\nu}$, and $\theta_{{\text{v}}_w} =\theta_{\text{wp}}$. $\theta_{{\text{v}}_1}$ was calculated as the volumetric moisture content corresponding to a pressure head ($\psi$) of 0.1 m~\cite{capraro2008neural}. The parameters $Y_m$ and $k_y$ in Equation~\ref{eq:yield_eqn} were set to 0.88 kg m$^{-2}$ and 1.15, respectively. These values were obtained from Reference~\cite{bennett2011crop} and represent calibrated values specific to the study area

\section{Uncontrolled inputs for the simulation period}

\subsection{Temperature and reference evapotranspiration}\label{section:weather_data}
\begin{figure}[H]
	\centerline{\includegraphics[width=0.60\columnwidth]{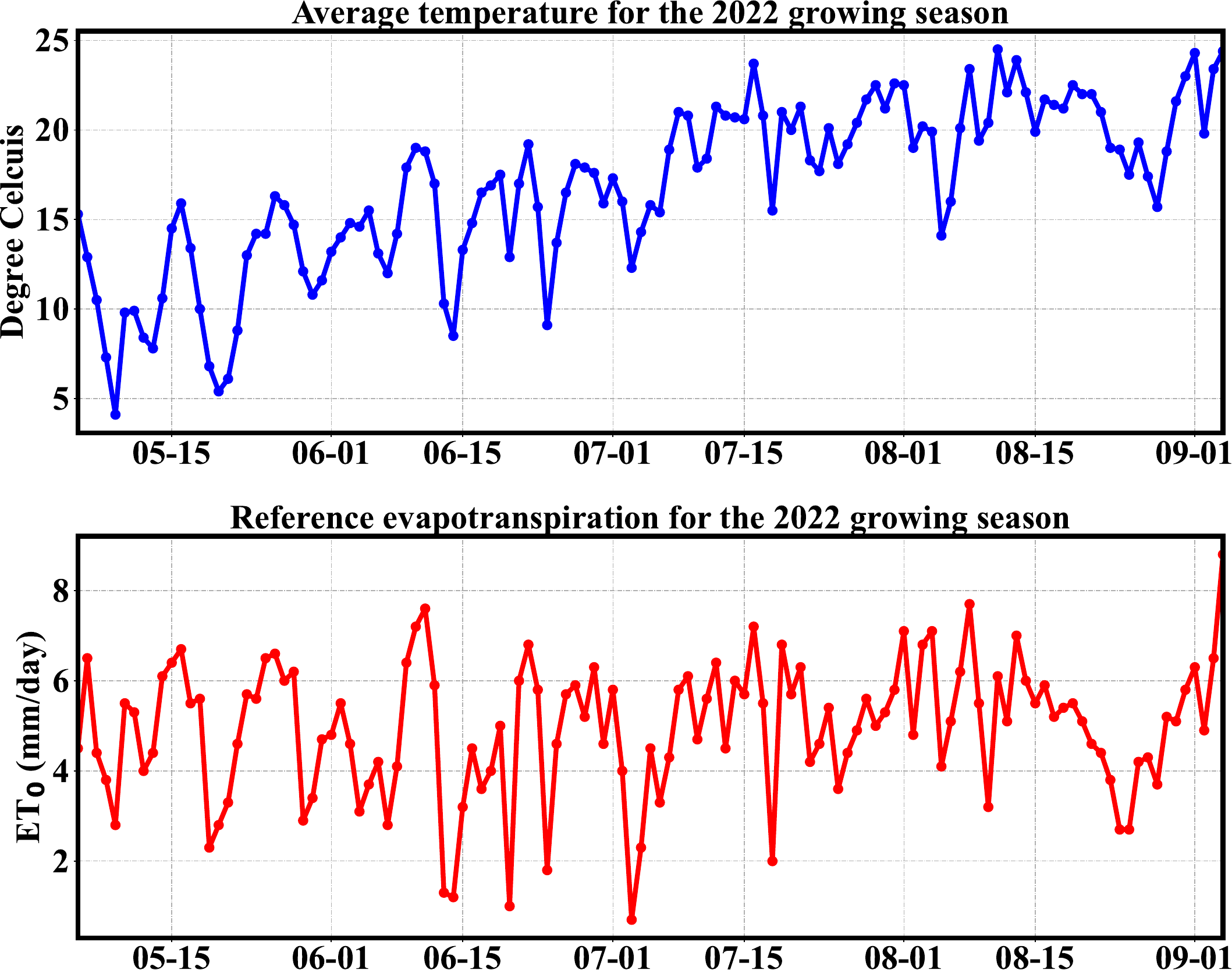}}
	\caption{} 
	\label{fig:weather_data}
\end{figure}

\subsection{Crop coefficient calculation}\label{sec:crop_coeff}
The coefficient ($\text{K}_{\text{c}}$)  of soft spring wheat was calculated as follows~\cite{bennett2011crop}:
\begin{equation}
	\label{eq:kc_relation}
	\text{K}_{\text{c}}(g) = -0.0207 + 0.00266g + \left(4.7\times10^{-8}\right)g^2 - \left(2.0\times10^{-9}\right)g^3 + \left(2.70\times10^{-13}\right)g^4
\end{equation} 
where $g$ is the cumulative growing-degree days (GDD). GDD is calculated as follows:
\begin{equation}
	\text{GDD} = \text{T}_{\text{avg}} - \text{T}_{\text{base}}
\end{equation}
where $\text{T}_{\text{avg}}$ is the daily average/mean temperature and $\text{T}_{\text{base}}$ is the base temperature below which crop growth ceases (5\textdegree C). 

\end{appendices}

\end{document}